\documentclass[sigconf,nonacm]{acmart}
\AtBeginDocument{
  \providecommand\BibTeX{{
    \normalfont B\kern-0.5em{\scshape i\kern-0.25em b}\kern-0.8em\TeX}}}

\usepackage{booktabs}
\usepackage{subcaption}
\usepackage{enumitem}
\usepackage{amsmath}
\usepackage{cleveref}
\usepackage{multirow}
\usepackage{url}
\usepackage{graphicx}
\usepackage{booktabs}
\usepackage{makecell}
\usepackage{array}
\usepackage{threeparttable}
\usepackage{pdflscape}
\usepackage[table]{xcolor}
\usepackage[most]{tcolorbox}

\newcounter{finding}
\newcommand{\finding}[1]{
    \refstepcounter{finding}
    \begin{tcolorbox}[
        enhanced,
        breakable, 
        sharp corners,
        boxrule=0pt, 
        leftrule=3pt, 
        colback=gray!15, 
        colframe=gray, 
        left=3pt, 
        right=3pt,
        top=2pt,
        bottom=2pt,
        arc=12pt, 
        before skip=1.5mm,
        after skip=1.5mm,
        width=\linewidth 
    ]
    \textbf{Ans. to RQ\arabic{finding}:} #1
    \end{tcolorbox}
}

\newcommand{\sd}[2]{#1}
\newcommand{\sdb}[2]{\textbf{#1}}

\author{Xinyue Li}
\email{xinyueli@stu.pku.edu.cn}
\affiliation{\institution{Peking University}\country{China}}

\author{Sixuan Li}
\email{lisixuan@stu.xidian.edu.cn}
\affiliation{\institution{Xidian University}\country{China}}

\author{Ying Xiao}
\email{ying.1.xiao@kcl.ac.uk}
\affiliation{\institution{King's College London}\country{United Kingdom}}

\author{Jie M. Zhang}
\email{jie.zhang@kcl.ac.uk}
\affiliation{\institution{King's College London}\country{United Kingdom}}

\author{Zhou Yang}
\email{zy25@ualberta.ca}
\affiliation{\institution{University of Alberta}\country{Canada}}

\author{Xuanzhe Liu}
\email{liuxuanzhe@pku.edu.cn}
\affiliation{\institution{Peking University}\country{China}}

\author{Zhenpeng Chen}
\authornote{Corresponding author.}
\email{zpchen@tsinghua.edu.cn}
\affiliation{\institution{Tsinghua University}\country{China}}

\begin{document}

\title{LLMs Are Not a Silver Bullet: A Case Study on Software Fairness}

\begin{abstract}
Fairness is a critical requirement for human-related, high-stakes software systems, motivating extensive research on bias mitigation. Prior work has largely focused on tabular data settings using traditional Machine Learning (ML) methods. With the rapid rise of Large Language Models (LLMs), recent studies have begun to explore their use for bias mitigation in the same setting. However, it remains unclear whether LLM-based methods offer advantages over traditional ML methods, leaving software engineers without clear guidance for practical adoption.
To address this gap, we present a large-scale study comparing state-of-the-art ML- and LLM-based bias mitigation methods. We find that ML-based methods consistently outperform LLM-based methods in both fairness and predictive performance, with even strong LLMs failing to surpass established ML baselines. To understand why prior LLM-based studies report favorable results, we analyze their evaluation settings and show that these gains are largely driven by artificially balanced test data rather than realistic imbalanced distributions. We further observe that existing LLM-based methods primarily rely on in-context learning and thus fail to leverage all available training data. Motivated by this, we explore supervised fine-tuning on the full training set and find that, while it achieves competitive results, its advantages over traditional ML methods remain limited. These findings suggest that LLMs are not a silver bullet for software fairness.

\end{abstract}



\maketitle

\section{Introduction}
Software fairness has become a critical requirement~\cite{aydemir2018roadmap, chen2024fairness, brun2018software, alidoosti2021ethics} for human-related, high-stakes software systems across socially sensitive domains, such as credit assessment~\cite{credit_dataset}, hiring~\cite{mahmoud2019performance}, and criminal justice~\cite{donohue2018replacement}. It requires software to provide equal opportunities or achieve comparable predictive performance across social groups defined by sensitive attributes such as sex, race, and age~\cite{chen2024fairness, brun2018software}. Violations of software fairness can lead to severe societal and legal consequences~\cite{wick2019unlocking, protection2018general}.

From the Software Engineering (SE) perspective, unfairness issues are commonly regarded as fairness bugs~ \cite{chen2024fairness, brun2018software}. Addressing these bugs has therefore become an important responsibility for software researchers and engineers~\cite{chen2024fairness}, motivating a growing body of work on bias mitigation in the SE community~\cite{chen2023comprehensive}. This body of work has largely focused on tabular data, where sensitive attributes are explicitly defined, fairness-sensitive applications are prevalent, and mitigation effects can be systematically evaluated. As a result, tabular data has become the most established and widely adopted setting for software fairness research~\cite{hort2024bias, soremekun2025software, biswas2020machine}.

Within this setting, prior bias mitigation research has been dominated by traditional Machine Learning (ML) methods~\cite{chakraborty2020fairway,mehrabi2021survey,hort2024bias}. Existing ML-based methods are typically categorized into pre-processing, in-processing, and post-processing methods, which mitigate bias by modifying the training data, incorporating fairness constraints during model training, or adjusting model outputs after training, respectively~\cite{hort2024bias, chen2023comprehensive}. These methods have been extensively studied and shown effective across diverse tabular datasets, establishing ML-based mitigation as the dominant paradigm for software fairness~\cite{chen2024fairness, chen2025software, chakraborty2020fairway, fairsense}.

Recently, with the rapid rise of Large Language Models (LLMs), software engineers are increasingly adopting LLMs as a general-purpose solution for a wide range of software tasks, often replacing traditional ML pipelines with LLM-based alternatives~\cite{abs240902977}. This trend has naturally extended to fairness-sensitive tabular prediction settings, where a growing body of work has begun to explore LLM-based bias mitigation in the same tabular setting~\cite{hu2024strategic,wang2023decodingtrust, liu2024confronting, wies2023learnability, cherepanova2025improving}.
These methods typically convert tabular data into natural language descriptions and use in-context learning, where a small set of demonstrations selected from the training data is included in the prompt alongside the test instances. The LLM then makes predictions conditioned on these demonstrations without updating its parameters. Prior studies~\cite{hu2024strategic, liu2024confronting, wang2023decodingtrust} report that carefully selected demonstrations can improve both predictive performance and fairness, suggesting that LLM-based methods may offer a promising alternative to traditional ML-based bias mitigation.

However, it remains unclear whether LLM-based methods offer advantages over traditional ML methods, leaving software engineers without clear guidance for practical adoption. This gap is partly due to a lack of direct comparisons between the two paradigms. For instance, recent LLM-based studies~\cite{hu2024strategic, liu2024confronting, cherepanova2025improving} propose in-context learning approaches for bias mitigation but do not compare against traditional ML-based methods. Similarly, recent work on ML-based bias mitigation~\cite{xiao2024mirrorfair, diversityICSE, chen2025software} rarely considers LLM-based approaches.

To address this gap, we conduct a large-scale empirical study comparing state-of-the-art ML- and LLM-based bias mitigation methods on widely used real-world tabular datasets. Specifically, we evaluate eight representative ML-based methods across four common ML models, and eight LLM-based methods under a unified experimental setting. Our results show that LLM-based methods are generally less effective than traditional ML-based approaches. On average, ML-based methods achieve both better fairness (48.3\%–59.6\% improvement across different fairness metrics) and higher predictive performance (e.g., 11.1\% higher accuracy and 9.3\% higher F1-score). This trend persists even when stronger LLMs are used.

However, existing LLM-based bias mitigation studies often report promising results, motivating us to investigate this discrepancy. We find that these studies~\cite{wang2023decodingtrust, hu2024strategic, liu2024confronting} typically evaluate on artificially balanced test sets with equal proportions across demographic groups and labels, whereas we follow real-world imbalanced distributions. To examine the impact of this difference, we compare LLM-based methods under both settings and observe that balanced distributions can improve fairness by 36.2\%–138.4\%. Such improvements, however, may not generalize to real-world settings, and thus provide limited guidance for software engineers in practice.

Another potential explanation for the performance gap between LLM- and ML-based methods is that existing LLM-based methods~\cite{hu2024strategic, wang2023decodingtrust} primarily rely on in-context learning, which uses only a small subset of the training data as demonstrations. To examine whether broader access to training data can improve LLM-based methods, we investigate supervised fine-tuning on the full training set. We consider both standard fine-tuning and fine-tuning combined with traditional data-level pre-processing. Our results show that fine-tuning substantially improves over in-context learning, but its advantages over traditional ML-based methods remain limited: no significant gains are observed in any fairness evaluations, and gains are observed in only 37.5\% of predictive performance comparisons.

Overall, this paper makes the following contributions:

\begin{itemize}[leftmargin=*]
\item A large-scale empirical study comparing state-of-the-art ML- and LLM-based bias mitigation methods on real-world tabular datasets, demonstrating that ML-based methods consistently outperform LLM-based methods.
\item An in-depth analysis of LLM-based methods, showing that their effectiveness is sensitive to evaluation settings and learning paradigms: balanced test distributions can inflate fairness, while fine-tuning still yields limited gains over traditional ML methods.
\item An open-source replication package~\cite{githublink}, including all scripts and data used in this study, to facilitate future research.
\end{itemize}

\section{Related Work}
This section summarizes existing work highly relevant to this study.

\noindent \textbf{Software Fairness.} Fairness has become an important requirement for modern software systems, motivating extensive research in the SE community. Existing software fairness research primarily focuses on tabular classification, the most widely studied setting in this area~\cite{mehrabi2021survey,chen2023comprehensive, hort2024bias}. Tabular classification supports decision-making across domains such as finance, healthcare, and criminal justice, where prediction outcomes directly affect individuals and social groups~\cite{chen2024fairness, aydemir2018roadmap, brun2018software}. Tabular datasets typically include sensitive attributes (e.g., sex, race, and age) that define privileged and unprivileged groups~\cite{mehrabi2021survey}. Because historical data may encode societal biases, learned models can produce systematically different outcomes across demographic groups, leading to fairness violations in intelligent software systems~\cite{soremekun2025software, li2026fairness}. Ensuring fairness in this setting is therefore essential for ethical decision-making.

\noindent \textbf{Traditional ML-Based Bias Mitigation.}
There have been extensive bias mitigation methods based on traditional ML methods~\cite{hort2024bias}.
These methods are commonly categorized into three types: pre-processing, in-processing, and post-processing. \emph{Pre-processing methods} modify the training data before learning, for example through reweighting, resampling, and synthetic data generation~\cite{peng2022fairmask, cot, ltdd, kamiran2012data}. LTDD~\cite{ltdd}, for instance, uses linear regression to identify non-sensitive features and feature values that are strongly associated with sensitive attributes, and excludes the biased parts while preserving as much unbiased information as possible. \emph{In-processing methods} mitigate bias during model training by modifying the learning objective or training procedure~\cite{zhang2018mitigating}. \emph{Post-processing methods} mitigate bias after training by adjusting prediction outputs~\cite{eop, roc}. Prior ML-based methods have also explored ensemble techniques~\cite{chen2022maat,xiao2024mirrorfair} that combine multiple mitigation strategies or models to leverage complementary strengths.

\noindent \textbf{Emerging LLM-Based Bias Mitigation.}
With the rapid advancement of LLMs, recent work has begun to explore their use for bias mitigation in tabular classification tasks~\cite{liu2024confronting, wang2023decodingtrust}. In this setting, tabular instances are typically converted into natural language descriptions and provided to LLMs for prediction. LLM-based methods usually operate through in-context learning, where the model conditions on task instructions and a small set of labeled examples (namely \emph{demonstrations}) included in the prompt. Unlike traditional ML-based methods, which intervene on training data, learning objectives, or model outputs, existing LLM-based bias mitigation methods primarily influence fairness through prompt design and demonstration selection.

LLM-based bias mitigation methods can be broadly categorized based on how demonstrations are constructed and used. \emph{Zero-shot} classification uses only task instructions and the input instance, serving as a non-mitigated baseline that reflects the model's inherent behavior~\cite{shaikh2023second,wang2023decodingtrust}. In contrast, \emph{few-shot} strategies mitigate bias by controlling the demonstrations included in the prompt. Prior work~\cite{liu2024confronting,wang2023decodingtrust} shows that fairness outcomes are highly sensitive to the composition and distribution of demonstrations. Existing methods therefore mainly improve fairness by balancing demographic groups and labels in the prompt, modifying demonstration labels, or selecting fairness-aware demonstration sets~\cite{wang2023decodingtrust,liu2024confronting}. More advanced methods search for fairness-aware demonstration sets that better balance fairness and predictive performance. For example, \emph{Fairness via Clustering-Genetic (FCG)}~\cite{hu2024strategic} clusters training data to identify representative samples and applies evolutionary search to optimize demonstration selection. 

Despite advances in both ML- and LLM-based bias mitigation methods, direct comparisons between the two paradigms remain limited. Existing studies typically evaluate methods within a single paradigm rather than across paradigms. For example, recent LLM-based work~\cite{hu2024strategic} compares different in-context learning strategies for bias mitigation but does not benchmark them against traditional ML-based approaches. As a result, it remains unclear how these two paradigms compare, leaving software engineers without clear guidance for practical adoption. To address this gap, this paper presents a large-scale empirical study comparing state-of-the-art methods from both paradigms.

\section{Experimental Setup}
This section describes the evaluation settings for this study.

\subsection{Datasets}

We evaluate bias mitigation methods on six tasks derived from three widely adopted real-world tabular datasets. 

Table~\ref{tab:datasets} summarizes the three datasets used in our study, include Adult~\cite{adult_dataset}, Compas~\cite{compas_dataset}, and Credit~\cite{credit_dataset}. All of them are well-established benchmarks in prior fairness research~\cite{chen2022maat,xiao2024mirrorfair, peng2022fairmask, hort2024bias}. They consist of real-world data collected from fairness-critical decision domains, including income prediction, recidivism risk assessment, and credit scoring. They also cover the three most commonly studied sensitive attributes in the literature, namely sex, race, and age~\cite{hort2024bias}. Based on the sensitive attributes available in each dataset, we define six fairness evaluation tasks, where each task corresponds to a specific dataset paired with a designated sensitive attribute. Specifically, the tasks include \textit{Adult-Sex}, \textit{Adult-Race}, \textit{Compas-Sex}, \textit{Compas-Race}, \textit{Credit-Sex}, and \textit{Credit-Age}.

\begin{table*}[h]
\centering
\caption{Benchmark datasets.}
\label{tab:datasets}
\small
\begin{tabular}{lllll}
\hline
\textbf{Name} & \textbf{Size} & \textbf{Sensitive attr(s)} & \textbf{Favorable label} & \textbf{Description} \\ \hline
Adult~\cite{adult_dataset} & 48,842 & sex, race & income $>$ 50K & Predict whether an individual’s income exceeds 50K. \\
Compas~\cite{compas_dataset} & 6,172 & sex, race & no recidivism & Predicting criminal defendant recidivism. \\
Credit~\cite{credit_dataset} & 30,000 & sex, age & no default & Predicting whether a customer will default on payment. \\ \hline
\end{tabular}
\end{table*}

\subsection{ML-Based Bias Mitigation Methods}

We evaluate one default baseline and eight traditional ML-based bias mitigation methods. We select representative and well-performing methods across all intervention stages, including pre-processing, in-processing, post-processing, and ensemble methods. Our selection encompasses both widely adopted methods in prior top-tier SE studies~\cite{mehrabi2021survey, hort2024bias, chen2025software, eop, peng2022fairmask, zhang2018mitigating, roc} and recently proposed state-of-the-art methods~\cite{xiao2024mirrorfair, ltdd, cot}, ensuring a comprehensive assessment that covers both classic foundations and the latest advancements in the field.

\begin{itemize}[leftmargin=*]
\item \textbf{Default baseline.} This configuration uses the original training data and the default ML pipeline without any bias mitigation.
\item \textbf{Pre-processing methods.} These methods mitigate bias by modifying the training data before model training. (i) \textbf{FairMask}~\cite{peng2022fairmask} uses an extrapolation model to adjust protected-attribute information before classification, thereby mitigating potential discrimination; (ii) \textbf{LTDD} (Linear-regression based Training Data Debugging)~\cite{ltdd} mitigates bias by identifying and excluding ``biased components'' of features, thereby cleaning the training data to support fairer predictions; (iii) \textbf{CoT} (Correlation Tuning)~\cite{cot} is a statistical intervention that adjusts data correlations via the Phi-coefficient and decouples sensitive attributes from the decision-making process to improve the fairness performance trade-off.
\item \textbf{In-processing methods.} These methods mitigate bias by incorporating fairness-aware constraints into the model training objective. \textbf{ADV}~\cite{zhang2018mitigating} applies adversarial learning to reduce the dependence between model predictions and sensitive attributes during training. It jointly optimizes a predictor and an adversary, discouraging discriminatory signals while preserving predictive performance.
\item \textbf{Post-processing methods.} These methods mitigate bias by adjusting model predictions after training without modifying the model. (i) \textbf{EOP}~\cite{eop} modifies output labels to equalize error rates across demographic groups, typically by aligning false positive and false negative rates; (ii) \textbf{ROC}~\cite{roc} adjusts predictions for instances near the decision boundary, altering labels in uncertain regions to improve fairness while preserving overall accuracy.
\item \textbf{Ensemble methods.} These methods combine multiple models or mitigation stages to balance fairness and predictive performance. (i) \textbf{MAAT}~\cite{chen2022maat} trains separate models optimized for predictive performance and fairness, and combines their predictions to balance the two objectives; (ii) \textbf{MirrorFair}~\cite{xiao2024mirrorfair} constructs a counterfactual dataset by flipping sensitive attributes, trains models on both the original and counterfactual data, and adaptively combines their predictions to produce fairer decisions.
\end{itemize}

\noindent\textbf{Model Selection.} For the traditional ML paradigm, we select four models widely studied in fairness research~\cite{hort2024bias, chen2025software, chen2022maat, xiao2024mirrorfair, diversityICSE} for training: Logistic Regression (LR), Random Forest (RF), Support Vector Machine (SVM), and Deep Neural Network (DNN). For the DNN architecture, we utilize a fully connected network with five hidden layers consisting of 64, 32, 16, 8, and 4 units, respectively.

\subsection{LLM-Based Bias Mitigation Methods} \label{methods}
We evaluate zero-shot prompting as the default baseline, along with eight LLM-based mitigation methods via in-context learning (ICL). We select these methods because they are representative prompt-level strategies studied in prior LLM fairness literature~\cite{liu2024confronting, wang2023decodingtrust, hu2024strategic} and cover the major intervention types in ICL-based bias mitigation, including baseline prompting, random demonstration selection, label manipulation, distribution-controlled demonstration construction, and fairness-aware demonstration search.

Let $k$ denote the number of demonstrations included in a prompt.
Let $A \in \{0,1\}$ denote the sensitive attribute (1: privileged, 0: unprivileged) and $Y \in \{0,1\}$ denote the original class label, where $Y=1$ represents the favorable label. 
For a given prompt, we define $r_A$ as the proportion of demonstrations drawn from the unprivileged group ($A=0$), and $r_Y$ as the proportion of demonstrations assigned the favorable label ($Y=1$).

\begin{itemize}[leftmargin=*]

\item \textbf{Zero Shot~\cite{liu2024confronting,wang2023decodingtrust}.} The LLM performs classification based solely on task instructions and the target instance, without demonstrations. This reflects the model's inherent behavior without contextual calibration.

\item \textbf{Random~\cite{liu2024confronting, wang2023decodingtrust}.} A fixed set of $k$ labeled demonstrations is randomly sampled from the training data and included in the prompt.

\item \textbf{Label Flipping~\cite{liu2024confronting}.} An intentional contextual intervention in which the labels of selected demonstrations are systematically flipped to weaken the empirical correlation between sensitive attributes and favorable outcomes. For example, a demonstration originally labeled as $(A=0, Y=0)$ is modified to $(A=0, Y=1)$, therefore increasing the representation of favorable outcomes for the unprivileged group within the prompt.

\item \textbf{Distribution-controlled Few-Shot~\cite{wang2023decodingtrust, hu2024strategic}.} Demonstrations are selected under three distinct distribution (S1, S2, S3) to analyze the impact of group and label representation. 

(i) \textbf{balanced (S1)}, where $r_A = 0.5$ and $r_Y = 0.5$, meaning half of the demonstrations are drawn from the unprivileged group and half are assigned the favorable label; 

(ii) \textbf{minority-balanced (S2)}, where $r_A = 1.0$ and $r_Y = 0.5$, meaning all demonstrations are drawn from the unprivileged group and half are assigned the favorable label;

(iii) \textbf{minority-unbalanced (S3)}, where $r_A = 1.0$ and $r_Y = 1.0$, meaning all demonstrations are drawn from the unprivileged group and all are assigned the favorable label.

\item \textbf{FCG Variants~\cite{hu2024strategic}.} Following the FCG strategy proposed in previous literature, candidate demonstration subsets are first constructed from clustered training samples and then optimized with a genetic search procedure based on both fairness and predictive performance. From the resulting optimized candidate pool $\mathcal{P}$, we select the final $k$-shot demonstration sets that satisfy the corresponding distribution requirements (S1, S2, S3) defined in the Few-Shot configurations. This gives three variants in our evaluation: \textbf{FCG-S1}, \textbf{FCG-S2}, and \textbf{FCG-S3}.

\end{itemize}

\noindent\textbf{Model Selection.} We use GPT-4o-mini~\cite{gpt4omini} as the primary LLM to implement these methods. We select GPT-4o-mini because it is a cost-effective model from a major AI vendor and has been widely used in prior LLM evaluation studies~\cite{chen2025more, gorti2024unboxing, liu2025llms}. Its relatively low inference cost also makes it suitable for our large-scale evaluation across multiple mitigation methods, six fairness evaluation tasks, and repeated runs.

\subsection{Evaluation Metrics}
Bias mitigation methods often involve a trade-off between fairness and predictive performance~\cite{chen2025software, hort2024bias, chen2022maat}; therefore, we evaluate their effectiveness using both fairness and performance metrics.

\subsubsection{Fairness Metrics}
We adopt group fairness metrics widely utilized in the previous literature to measure software bias~\cite{mehrabi2021survey, chen2022maat, biswas2020machine, biswas2021fair, chakraborty2021bias, zhang2021ignorance}. As defined in Section~\ref{methods}, let $A$ be a protected attribute, with $1$ representing the privileged group and $0$ the unprivileged group. Let $Y$ be the original class label and $\hat{Y}$ the predicted label, where $1$ denotes the favorable class and $0$ the unfavorable class; let $P$ denote the probability. Following previous work~\cite{chen2022maat, hort2024bias, xiao2024mirrorfair, mehrabi2021survey, diversityICSE}, we use the absolute values of these metrics, where a value of $0$ indicates perfect fairness and larger values indicate higher levels of bias.

\begin{itemize}[leftmargin=*]
    \item \textbf{Statistical Parity Difference (SPD)} measures the disparity in the probability of receiving a favorable prediction between unprivileged and privileged groups:
    \begin{equation}
    \small
        SPD = |P(\hat{Y}=1 | A=0) - P(\hat{Y}=1 | A=1)|
    \end{equation}

    \item \textbf{Average Odds Difference (AOD)} evaluates the average of the absolute differences in false-positive rates and true-positive rates between the two demographic groups:
    \begin{equation}
    \small
    \begin{aligned}[c]
        AOD = \frac{1}{2} ( & |P(\hat{Y}=1|A=0, Y=0) - P(\hat{Y}=1|A=1, Y=0)| \\
        + & |P(\hat{Y}=1|A=0, Y=1) - P(\hat{Y}=1|A=1, Y=1)| )
    \end{aligned}
    \end{equation}

    \item \textbf{Equal Opportunity Difference (EOD)} measures the true-positive rate difference between unprivileged and privileged groups:
    \begin{equation}
    \small
        EOD = |P(\hat{Y}=1 | A=0, Y=1) - P(\hat{Y}=1 | A=1, Y=1)|
    \end{equation}
\end{itemize}

\subsubsection{Performance Metrics}
To measure predictive performance, we utilize traditional classification metrics, including precision (Prec.), recall (Rec.), F1-score (F1), and accuracy (Acc.). For a given class, precision is defined as the proportion of samples predicted as that class that actually belong to it, while recall denotes the proportion of samples belonging to a class that are correctly predicted. F1-score is calculated as the harmonic mean of precision and recall. Following established practices in SE research~\cite{hort2024bias, zhang2021ignorance, diversityICSE, chen2022maat}, we report the macro-average values for precision, recall, and F1-score to enable a balanced comparison of the overall performance across both favorable and unfavorable classes. This involves calculating each metric for each class individually and then averaging the results. Accuracy, which measures the frequency of correct predictions, remains a standard metric in fairness literature~\cite{chen2022maat, diversityICSE, chen2024fairness}. For all four metrics, larger values indicate superior predictive performance, with 1 representing perfect prediction.

\subsection{Implementation Details}\label{sec:implementation}

\noindent We briefly describe the implementation details used in our study.

\noindent\textbf{Experimental environment.} All experiments are implemented in Python 3.11.9. For the traditional ML paradigm, we use IBM AIF360~\cite{aif360} for mitigation methods and fairness metrics, \textit{Scikit-learn} for traditional classifiers, and \textit{TensorFlow Keras} for the DNN model. 
LLM inference is performed via API calls~\cite{openai, openrouter}, using the default temperature for each evaluated model to reflect real-world application scenarios, where users typically interact with these models in their out-of-the-box configurations. Fine-tuning experiments are conducted on a machine with one NVIDIA RTX 5090 GPU (32GB) and 90GB RAM.

\noindent\textbf{Data preprocessing and split.} Before partitioning the datasets, we perform a standard preprocessing step by removing all instances containing missing values (NA) to ensure data quality and consistency across different models. To mitigate the impact of randomness and ensure the soundness of our comparisons, we then adopt a consistent evaluation protocol across both traditional ML and LLM paradigms. In each experimental run, the cleaned dataset is randomly split into 80\% training data and 20\% testing data using different random seeds. We do not employ cross-validation to maintain alignment with the standard prompting and fine-tuning procedures used in recent LLM studies. Following established practices in software fairness literature~\cite{chen2022maat, xiao2024mirrorfair, diversityICSE} to reduce random variation, each bias mitigation task is repeated 3 times with different random seeds, and we report the average predictive performance and fairness metrics across these runs.

\begin{figure}[t]
    \centering
    \includegraphics[width=0.4\textwidth]{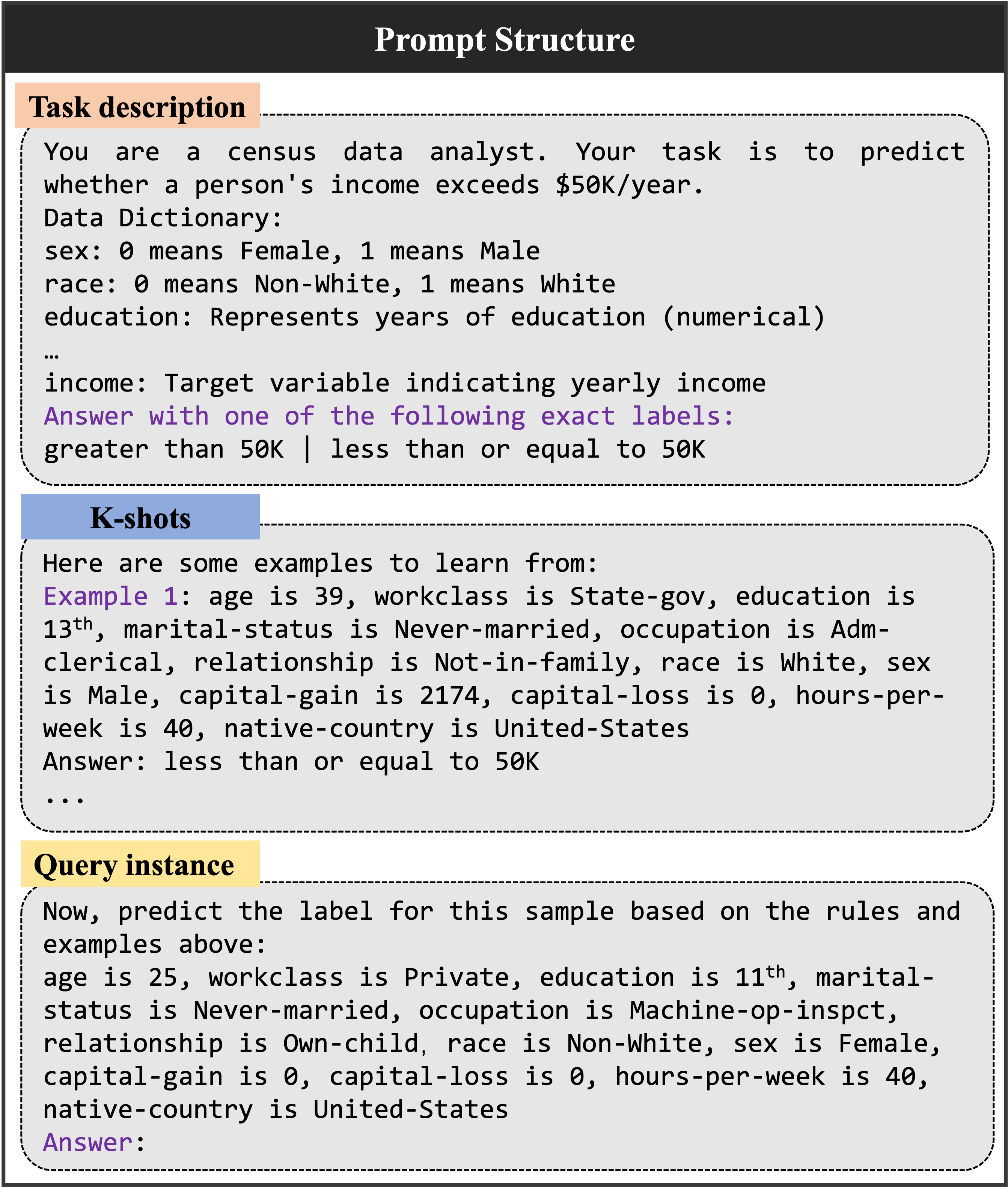}
    \caption{Prompt structure for the LLM paradigm.}
    \label{fig:prompt}
\end{figure}

\noindent\textbf{Prompt serialization.} To facilitate LLM processing of tabular data, we serialize each structured record into a natural language string. As in prior studies~\cite{liu2024confronting, hu2024strategic}, each instance is represented as a concatenation of feature-value pairs (e.g., ``$f_1$ is $x_1, \dots, f_d$ is $x_d$''). As illustrated in Figure~\ref{fig:prompt}, each prompt consists of a task description, $k$-shot demonstrations, and a query instance. Following previous studies~\cite{wang2023decodingtrust, hu2024strategic, chen2025software}, we set $k=16$; in the zero-shot setting, we use the same prompt format without demonstrations.

\section{Research Questions and Results}
This section presents our research questions (RQs) and answers them through experimental results.

\subsection{RQ1: How do LLM-based and ML-based bias mitigation methods compare?}

\subsubsection{Motivation and Methodology.} In this RQ, we compare the overall effectiveness of the traditional ML and LLM paradigms for tabular bias mitigation across six fairness evaluation tasks. Table~\ref{tab:rq1_overall} reports the detailed results. Each value is averaged over three repeated runs with different random train-test splits. For traditional ML, each method result is further averaged across four classification models (i.e., SVM, RF, LR, and DNN), while all results for the LLM paradigm are obtained using GPT-4o-mini~\cite{gpt4omini}. In Table~\ref{tab:rq1_overall}, \textit{Default} and \textit{Zero-Shot} denote the baseline settings for the ML and LLM paradigms, respectively, and \textit{Average} denotes the mean result of the eight bias mitigation methods under each paradigm.

To assess whether the predictive-performance and fairness differences between the two paradigms are statistically significant, we conduct the non-parametric Mann-Whitney U-test~\cite{mann1947test} on the raw outcomes. For each task, we compare the raw outcomes of the two paradigms on the four predictive metrics reported in Table~\ref{tab:rq1_overall} (Accuracy, Recall, Precision, and F1), resulting in 24 predictive-performance comparisons ($24 = 6$ tasks $\times 4$ metrics), and on the three fairness metrics (SPD, EOD, and AOD), resulting in 18 fairness comparisons ($18 = 6$ tasks $\times 3$ metrics). Both groups aggregate raw results across eight mitigation methods and repeated runs, with the ML group further incorporating the corresponding base models. A difference is considered statistically significant when the resulting $p$-value is below 0.05. Since higher values indicate better predictive performance, a predictive case is considered to favor ML when the ML group has a higher mean value and the Mann-Whitney test produces $p<0.05$. Since lower values indicate better fairness, a fairness case is considered to favor ML when the ML group has a lower mean value and the Mann-Whitney test produces $p<0.05$.

\subsubsection{Results.}

\begin{table*}[t]
\centering
\scriptsize
\renewcommand{\arraystretch}{0.95}
\caption{(RQ1) Predictive performance and fairness of ML- and LLM-based bias mitigation methods across six tasks. Each paradigm includes one baseline (\textit{Default} for ML and \textit{Zero-Shot} for LLM) and eight mitigation methods. Metrics labeled with $\uparrow$ indicate that higher values are better, while those labeled with $\downarrow$ indicate that lower values are better. Results for ML-based methods are averaged over SVM, RF, LR, and DNN. The \textit{Average} rows report the mean results of the eight mitigation methods within each paradigm.}
\label{tab:rq1_overall}
\setlength{\tabcolsep}{2pt}
\resizebox{\textwidth}{!}{
\begin{tabular}{l|l|lllllll|lllllll|lllllll}
\toprule
\multirow{2}{*}{Paradigm} & \multirow{2}{*}{Method}
& \multicolumn{7}{c|}{Adult-Sex}
& \multicolumn{7}{c|}{Compas-Sex}
& \multicolumn{7}{c}{Credit-Sex} \\
\cmidrule(lr){3-9}\cmidrule(lr){10-16}\cmidrule(lr){17-23}
& & Acc.$\uparrow$ & F1$\uparrow$ & Prec.$\uparrow$ & Rec.$\uparrow$ & SPD$\downarrow$ & EOD$\downarrow$ & AOD$\downarrow$
  & Acc.$\uparrow$ & F1$\uparrow$ & Prec.$\uparrow$ & Rec.$\uparrow$ & SPD$\downarrow$ & EOD$\downarrow$ & AOD$\downarrow$
  & Acc.$\uparrow$ & F1$\uparrow$ & Prec.$\uparrow$ & Rec.$\uparrow$ & SPD$\downarrow$ & EOD$\downarrow$ & AOD$\downarrow$ \\
\midrule
\multirow{10}{*}{ML}
& Default    & \sd{0.849}{0.004} & \sd{0.784}{0.006} & \sd{0.808}{0.010} & \sd{0.768}{0.005} & \sd{0.181}{0.006} & \sd{0.096}{0.028} & \sd{0.086}{0.018} & \sd{0.646}{0.020} & \sd{0.641}{0.022} & \sd{0.644}{0.022} & \sd{0.641}{0.022} & \sd{0.284}{0.032} & \sd{0.221}{0.026} & \sd{0.259}{0.033} & \sd{0.813}{0.006} & \sd{0.648}{0.010} & \sd{0.757}{0.014} & \sd{0.628}{0.007} & \sd{0.021}{0.007} & \sd{0.011}{0.001} & \sd{0.018}{0.009} \\
\cmidrule(lr){2-23}
& COT        & \sd{0.845}{0.003} & \sd{0.772}{0.004} & \sd{0.807}{0.008} & \sd{0.751}{0.002} & \sd{0.133}{0.008} & \sd{0.028}{0.020} & \sd{0.037}{0.007} & \sd{0.651}{0.021} & \sd{0.643}{0.021} & \sd{0.647}{0.022} & \sd{0.643}{0.021} & \sd{0.060}{0.022} & \sd{0.046}{0.019} & \sd{0.051}{0.020} & \sd{0.812}{0.006} & \sd{0.649}{0.010} & \sd{0.753}{0.011} & \sd{0.629}{0.008} & \sd{0.009}{0.005} & \sd{0.004}{0.001} & \sd{0.015}{0.006} \\
& LTDD       & \sd{0.844}{0.003} & \sd{0.775}{0.006} & \sd{0.801}{0.004} & \sd{0.758}{0.006} & \sd{0.133}{0.006} & \sd{0.120}{0.016} & \sd{0.082}{0.009} & \sd{0.652}{0.021} & \sd{0.643}{0.022} & \sd{0.648}{0.022} & \sd{0.643}{0.021} & \sd{0.073}{0.008} & \sd{0.078}{0.013} & \sd{0.089}{0.011} & \sd{0.813}{0.006} & \sd{0.647}{0.009} & \sd{0.756}{0.013} & \sd{0.628}{0.007} & \sd{0.010}{0.001} & \sd{0.005}{0.001} & \sd{0.016}{0.006} \\
& FairMask   & \sd{0.849}{0.004} & \sd{0.783}{0.006} & \sd{0.808}{0.010} & \sd{0.766}{0.005} & \sd{0.176}{0.005} & \sd{0.084}{0.024} & \sd{0.079}{0.015} & \sd{0.645}{0.020} & \sd{0.640}{0.021} & \sd{0.645}{0.021} & \sd{0.642}{0.021} & \sd{0.142}{0.016} & \sd{0.097}{0.032} & \sd{0.106}{0.016} & \sd{0.813}{0.006} & \sd{0.652}{0.009} & \sd{0.755}{0.013} & \sd{0.631}{0.007} & \sd{0.011}{0.007} & \sd{0.005}{0.001} & \sd{0.011}{0.002} \\
& ADV        & \sd{0.841}{0.006} & \sd{0.762}{0.014} & \sd{0.806}{0.011} & \sd{0.739}{0.018} & \sd{0.048}{0.036} & \sd{0.218}{0.067} & \sd{0.117}{0.043} & \sd{0.650}{0.010} & \sd{0.645}{0.013} & \sd{0.647}{0.013} & \sd{0.646}{0.015} & \sd{0.348}{0.220} & \sd{0.269}{0.178} & \sd{0.329}{0.230} & \sd{0.819}{0.005} & \sd{0.672}{0.007} & \sd{0.758}{0.015} & \sd{0.647}{0.006} & \sd{0.008}{0.010} & \sd{0.007}{0.006} & \sd{0.023}{0.019} \\
& MAAT       & \sd{0.844}{0.001} & \sd{0.758}{0.004} & \sd{0.823}{0.006} & \sd{0.730}{0.004} & \sd{0.102}{0.005} & \sd{0.060}{0.029} & \sd{0.043}{0.012} & \sd{0.651}{0.022} & \sd{0.641}{0.023} & \sd{0.648}{0.024} & \sd{0.641}{0.023} & \sd{0.149}{0.029} & \sd{0.090}{0.032} & \sd{0.119}{0.035} & \sd{0.813}{0.006} & \sd{0.645}{0.009} & \sd{0.757}{0.013} & \sd{0.625}{0.007} & \sd{0.010}{0.007} & \sd{0.006}{0.000} & \sd{0.013}{0.001} \\
& MirrorFair & \sd{0.847}{0.002} & \sd{0.772}{0.005} & \sd{0.814}{0.008} & \sd{0.749}{0.005} & \sd{0.130}{0.003} & \sd{0.043}{0.005} & \sd{0.043}{0.003} & \sd{0.648}{0.020} & \sd{0.635}{0.022} & \sd{0.647}{0.022} & \sd{0.636}{0.021} & \sd{0.145}{0.017} & \sd{0.090}{0.022} & \sd{0.115}{0.024} & \sd{0.813}{0.006} & \sd{0.648}{0.009} & \sd{0.757}{0.012} & \sd{0.628}{0.006} & \sd{0.010}{0.007} & \sd{0.005}{0.000} & \sd{0.011}{0.002} \\
& EOP        & \sd{0.826}{0.005} & \sd{0.752}{0.006} & \sd{0.772}{0.010} & \sd{0.739}{0.004} & \sd{0.105}{0.008} & \sd{0.037}{0.023} & \sd{0.026}{0.011} & \sd{0.609}{0.008} & \sd{0.589}{0.012} & \sd{0.605}{0.015} & \sd{0.593}{0.013} & \sd{0.040}{0.021} & \sd{0.025}{0.020} & \sd{0.028}{0.015} & \sd{0.809}{0.006} & \sd{0.641}{0.009} & \sd{0.744}{0.012} & \sd{0.623}{0.007} & \sd{0.010}{0.004} & \sd{0.007}{0.002} & \sd{0.014}{0.002} \\
& ROC        & \sd{0.776}{0.006} & \sd{0.737}{0.008} & \sd{0.726}{0.008} & \sd{0.781}{0.008} & \sd{0.036}{0.003} & \sd{0.157}{0.008} & \sd{0.118}{0.006} & \sd{0.647}{0.021} & \sd{0.642}{0.021} & \sd{0.645}{0.020} & \sd{0.643}{0.021} & \sd{0.057}{0.017} & \sd{0.042}{0.027} & \sd{0.046}{0.029} & \sd{0.787}{0.002} & \sd{0.689}{0.005} & \sd{0.692}{0.003} & \sd{0.689}{0.007} & \sd{0.027}{0.006} & \sd{0.018}{0.007} & \sd{0.017}{0.003} \\
\cmidrule(lr){2-23}
& \textbf{Average} & \sdb{0.834}{0.003} & \sdb{0.764}{0.005} & \sdb{0.795}{0.008} & \sdb{0.752}{0.004} & \sdb{0.108}{0.007} & \sdb{0.093}{0.005} & \sdb{0.068}{0.004}
& \sdb{0.644}{0.017} & \sdb{0.635}{0.018} & \sdb{0.642}{0.018} & \sdb{0.636}{0.018} & \sdb{0.127}{0.029} & \sdb{0.092}{0.020} & \sdb{0.111}{0.025}
& \sdb{0.810}{0.005} & \sdb{0.655}{0.008} & \sdb{0.746}{0.011} & \sdb{0.637}{0.006} & \sdb{0.012}{0.004} & \sdb{0.007}{0.002} & \sdb{0.015}{0.004} \\
\cmidrule(lr){1-23}
\multirow{10}{*}{LLM}
& Zero-Shot & \sd{0.768}{0.001} & \sd{0.728}{0.001} & \sd{0.717}{0.001} & \sd{0.765}{0.002} & \sd{0.262}{0.002} & \sd{0.179}{0.001} & \sd{0.169}{0.001} & \sd{0.642}{0.002} & \sd{0.641}{0.002} & \sd{0.661}{0.002} & \sd{0.655}{0.002} & \sd{0.236}{0.008} & \sd{0.149}{0.020} & \sd{0.182}{0.011} & \sd{0.779}{0.001} & \sd{0.557}{0.002} & \sd{0.644}{0.005} & \sd{0.557}{0.001} & \sd{0.087}{0.004} & \sd{0.194}{0.005} & \sd{0.124}{0.004} \\
\cmidrule(lr){2-23}
& Random    & \sd{0.775}{0.021} & \sd{0.717}{0.018} & \sd{0.710}{0.021} & \sd{0.734}{0.009} & \sd{0.145}{0.002} & \sd{0.079}{0.019} & \sd{0.065}{0.012} & \sd{0.591}{0.092} & \sd{0.580}{0.092} & \sd{0.584}{0.094} & \sd{0.581}{0.090} & \sd{0.213}{0.046} & \sd{0.196}{0.086} & \sd{0.193}{0.037} & \sd{0.644}{0.026} & \sd{0.593}{0.028} & \sd{0.605}{0.027} & \sd{0.648}{0.039} & \sd{0.016}{0.005} & \sd{0.028}{0.018} & \sd{0.019}{0.010} \\
& LF        & \sd{0.703}{0.015} & \sd{0.662}{0.007} & \sd{0.664}{0.011} & \sd{0.710}{0.027} & \sd{0.210}{0.073} & \sd{0.155}{0.032} & \sd{0.142}{0.052} & \sd{0.516}{0.113} & \sd{0.512}{0.114} & \sd{0.513}{0.113} & \sd{0.513}{0.114} & \sd{0.250}{0.066} & \sd{0.243}{0.063} & \sd{0.227}{0.081} & \sd{0.670}{0.073} & \sd{0.595}{0.027} & \sd{0.608}{0.023} & \sd{0.632}{0.025} & \sd{0.031}{0.007} & \sd{0.039}{0.008} & \sd{0.038}{0.009} \\
& S1        & \sd{0.780}{0.021} & \sd{0.719}{0.001} & \sd{0.715}{0.010} & \sd{0.732}{0.019} & \sd{0.156}{0.033} & \sd{0.096}{0.094} & \sd{0.080}{0.065} & \sd{0.640}{0.020} & \sd{0.631}{0.021} & \sd{0.638}{0.025} & \sd{0.632}{0.022} & \sd{0.015}{0.011} & \sd{0.026}{0.021} & \sd{0.028}{0.015} & \sd{0.711}{0.035} & \sd{0.617}{0.014} & \sd{0.615}{0.013} & \sd{0.634}{0.019} & \sd{0.025}{0.016} & \sd{0.033}{0.029} & \sd{0.030}{0.024} \\
& S2        & \sd{0.771}{0.028} & \sd{0.708}{0.004} & \sd{0.709}{0.019} & \sd{0.725}{0.029} & \sd{0.152}{0.017} & \sd{0.076}{0.037} & \sd{0.070}{0.018} & \sd{0.596}{0.066} & \sd{0.593}{0.065} & \sd{0.593}{0.065} & \sd{0.594}{0.066} & \sd{0.070}{0.073} & \sd{0.055}{0.090} & \sd{0.072}{0.051} & \sd{0.628}{0.078} & \sd{0.562}{0.029} & \sd{0.585}{0.017} & \sd{0.610}{0.027} & \sd{0.025}{0.025} & \sd{0.033}{0.023} & \sd{0.032}{0.025} \\
& S3        & \sd{0.797}{0.008} & \sd{0.727}{0.008} & \sd{0.729}{0.003} & \sd{0.726}{0.019} & \sd{0.128}{0.032} & \sd{0.057}{0.038} & \sd{0.048}{0.028} & \sd{0.634}{0.005} & \sd{0.627}{0.008} & \sd{0.631}{0.010} & \sd{0.629}{0.010} & \sd{0.183}{0.133} & \sd{0.158}{0.138} & \sd{0.207}{0.139} & \sd{0.713}{0.005} & \sd{0.575}{0.022} & \sd{0.577}{0.015} & \sd{0.576}{0.025} & \sd{0.080}{0.011} & \sd{0.096}{0.009} & \sd{0.089}{0.009} \\
& FCG-S1    & \sd{0.817}{0.012} & \sd{0.686}{0.055} & \sd{0.809}{0.010} & \sd{0.663}{0.053} & \sd{0.088}{0.034} & \sd{0.028}{0.015} & \sd{0.026}{0.012} & \sd{0.638}{0.035} & \sd{0.588}{0.052} & \sd{0.665}{0.025} & \sd{0.611}{0.034} & \sd{0.156}{0.078} & \sd{0.222}{0.108} & \sd{0.148}{0.077} & \sd{0.639}{0.042} & \sd{0.588}{0.022} & \sd{0.605}{0.006} & \sd{0.647}{0.016} & \sd{0.014}{0.003} & \sd{0.017}{0.007} & \sd{0.018}{0.005} \\
& FCG-S2    & \sd{0.804}{0.012} & \sd{0.737}{0.007} & \sd{0.740}{0.008} & \sd{0.739}{0.022} & \sd{0.220}{0.020} & \sd{0.196}{0.094} & \sd{0.158}{0.056} & \sd{0.651}{0.020} & \sd{0.613}{0.037} & \sd{0.672}{0.014} & \sd{0.627}{0.026} & \sd{0.105}{0.054} & \sd{0.093}{0.060} & \sd{0.076}{0.048} & \sd{0.660}{0.035} & \sd{0.606}{0.025} & \sd{0.614}{0.012} & \sd{0.658}{0.011} & \sd{0.018}{0.014} & \sd{0.021}{0.010} & \sd{0.024}{0.014} \\
& FCG-S3    & \sd{0.829}{0.007} & \sd{0.725}{0.027} & \sd{0.807}{0.018} & \sd{0.697}{0.028} & \sd{0.132}{0.020} & \sd{0.136}{0.061} & \sd{0.089}{0.035} & \sd{0.646}{0.023} & \sd{0.605}{0.052} & \sd{0.669}{0.017} & \sd{0.621}{0.035} & \sd{0.054}{0.010} & \sd{0.052}{0.049} & \sd{0.029}{0.024} & \sd{0.688}{0.013} & \sd{0.590}{0.012} & \sd{0.587}{0.011} & \sd{0.604}{0.022} & \sd{0.087}{0.041} & \sd{0.103}{0.079} & \sd{0.096}{0.055} \\
\cmidrule(lr){2-23}
& \textbf{Average} & \sdb{0.785}{0.003} & \sdb{0.710}{0.013} & \sdb{0.735}{0.002} & \sdb{0.716}{0.020} & \sdb{0.154}{0.013} & \sdb{0.103}{0.041} & \sdb{0.085}{0.026}
& \sdb{0.614}{0.023} & \sdb{0.594}{0.028} & \sdb{0.621}{0.018} & \sdb{0.601}{0.023} & \sdb{0.131}{0.032} & \sdb{0.131}{0.038} & \sdb{0.122}{0.033}
& \sdb{0.669}{0.018} & \sdb{0.591}{0.004} & \sdb{0.600}{0.002} & \sdb{0.626}{0.008} & \sdb{0.037}{0.001} & \sdb{0.046}{0.011} & \sdb{0.043}{0.004} \\
\midrule

\multirow{2}{*}{Paradigm} & \multirow{2}{*}{Method}
& \multicolumn{7}{c|}{Adult-Race}
& \multicolumn{7}{c|}{Compas-Race}
& \multicolumn{7}{c}{Credit-Age} \\
\cmidrule(lr){3-9}\cmidrule(lr){10-16}\cmidrule(lr){17-23}
& & Acc.$\uparrow$ & F1$\uparrow$ & Prec.$\uparrow$ & Rec.$\uparrow$ & SPD$\downarrow$ & EOD$\downarrow$ & AOD$\downarrow$
  & Acc.$\uparrow$ & F1$\uparrow$ & Prec.$\uparrow$ & Rec.$\uparrow$ & SPD$\downarrow$ & EOD$\downarrow$ & AOD$\downarrow$
  & Acc.$\uparrow$ & F1$\uparrow$ & Prec.$\uparrow$ & Rec.$\uparrow$ & SPD$\downarrow$ & EOD$\downarrow$ & AOD$\downarrow$ \\
\midrule
\multirow{10}{*}{ML}
& Default    & \sd{0.849}{0.004} & \sd{0.784}{0.006} & \sd{0.808}{0.010} & \sd{0.768}{0.005} & \sd{0.098}{0.006} & \sd{0.077}{0.026} & \sd{0.057}{0.014} & \sd{0.646}{0.020} & \sd{0.641}{0.022} & \sd{0.644}{0.022} & \sd{0.641}{0.022} & \sd{0.287}{0.022} & \sd{0.236}{0.027} & \sd{0.257}{0.025} & \sd{0.813}{0.006} & \sd{0.648}{0.010} & \sd{0.757}{0.014} & \sd{0.628}{0.007} & \sd{0.053}{0.009} & \sd{0.030}{0.015} & \sd{0.047}{0.004} \\
\cmidrule(lr){2-23}
& COT        & \sd{0.847}{0.004} & \sd{0.783}{0.006} & \sd{0.803}{0.010} & \sd{0.769}{0.004} & \sd{0.055}{0.004} & \sd{0.040}{0.015} & \sd{0.025}{0.010} & \sd{0.643}{0.026} & \sd{0.640}{0.026} & \sd{0.641}{0.026} & \sd{0.641}{0.025} & \sd{0.080}{0.052} & \sd{0.058}{0.032} & \sd{0.054}{0.034} & \sd{0.812}{0.005} & \sd{0.648}{0.009} & \sd{0.754}{0.010} & \sd{0.629}{0.007} & \sd{0.026}{0.005} & \sd{0.015}{0.008} & \sd{0.033}{0.002} \\
& LTDD       & \sd{0.848}{0.003} & \sd{0.783}{0.007} & \sd{0.805}{0.007} & \sd{0.767}{0.007} & \sd{0.057}{0.002} & \sd{0.099}{0.014} & \sd{0.065}{0.006} & \sd{0.650}{0.020} & \sd{0.643}{0.020} & \sd{0.647}{0.021} & \sd{0.643}{0.019} & \sd{0.097}{0.029} & \sd{0.115}{0.032} & \sd{0.114}{0.026} & \sd{0.812}{0.005} & \sd{0.645}{0.008} & \sd{0.755}{0.010} & \sd{0.626}{0.006} & \sd{0.029}{0.003} & \sd{0.014}{0.002} & \sd{0.044}{0.006} \\
& FairMask   & \sd{0.849}{0.004} & \sd{0.783}{0.006} & \sd{0.807}{0.010} & \sd{0.767}{0.005} & \sd{0.092}{0.009} & \sd{0.060}{0.031} & \sd{0.047}{0.017} & \sd{0.639}{0.024} & \sd{0.635}{0.026} & \sd{0.637}{0.027} & \sd{0.636}{0.026} & \sd{0.192}{0.030} & \sd{0.151}{0.037} & \sd{0.161}{0.027} & \sd{0.813}{0.006} & \sd{0.650}{0.009} & \sd{0.756}{0.014} & \sd{0.629}{0.007} & \sd{0.052}{0.008} & \sd{0.028}{0.017} & \sd{0.043}{0.002} \\
& ADV        & \sd{0.848}{0.004} & \sd{0.781}{0.004} & \sd{0.808}{0.015} & \sd{0.764}{0.008} & \sd{0.021}{0.013} & \sd{0.107}{0.046} & \sd{0.062}{0.026} & \sd{0.653}{0.012} & \sd{0.646}{0.013} & \sd{0.650}{0.010} & \sd{0.646}{0.013} & \sd{0.137}{0.069} & \sd{0.113}{0.115} & \sd{0.134}{0.116} & \sd{0.818}{0.007} & \sd{0.665}{0.012} & \sd{0.760}{0.011} & \sd{0.641}{0.010} & \sd{0.018}{0.014} & \sd{0.007}{0.006} & \sd{0.029}{0.021} \\
& MAAT       & \sd{0.847}{0.002} & \sd{0.771}{0.005} & \sd{0.817}{0.008} & \sd{0.746}{0.004} & \sd{0.062}{0.004} & \sd{0.021}{0.007} & \sd{0.017}{0.006} & \sd{0.644}{0.026} & \sd{0.638}{0.027} & \sd{0.640}{0.027} & \sd{0.638}{0.026} & \sd{0.161}{0.040} & \sd{0.112}{0.040} & \sd{0.129}{0.034} & \sd{0.812}{0.006} & \sd{0.644}{0.009} & \sd{0.756}{0.014} & \sd{0.626}{0.007} & \sd{0.039}{0.007} & \sd{0.021}{0.012} & \sd{0.029}{0.002} \\
& MirrorFair & \sd{0.850}{0.004} & \sd{0.783}{0.007} & \sd{0.811}{0.010} & \sd{0.765}{0.006} & \sd{0.088}{0.008} & \sd{0.054}{0.030} & \sd{0.042}{0.017} & \sd{0.644}{0.026} & \sd{0.636}{0.028} & \sd{0.641}{0.028} & \sd{0.636}{0.027} & \sd{0.202}{0.034} & \sd{0.147}{0.040} & \sd{0.173}{0.030} & \sd{0.813}{0.006} & \sd{0.650}{0.008} & \sd{0.756}{0.012} & \sd{0.629}{0.006} & \sd{0.048}{0.008} & \sd{0.025}{0.018} & \sd{0.038}{0.002} \\
& EOP        & \sd{0.841}{0.003} & \sd{0.769}{0.008} & \sd{0.797}{0.010} & \sd{0.752}{0.008} & \sd{0.061}{0.005} & \sd{0.046}{0.028} & \sd{0.027}{0.014} & \sd{0.605}{0.008} & \sd{0.594}{0.009} & \sd{0.600}{0.009} & \sd{0.595}{0.009} & \sd{0.024}{0.020} & \sd{0.022}{0.016} & \sd{0.030}{0.004} & \sd{0.804}{0.004} & \sd{0.642}{0.008} & \sd{0.722}{0.004} & \sd{0.624}{0.006} & \sd{0.027}{0.006} & \sd{0.018}{0.003} & \sd{0.027}{0.008} \\
& ROC        & \sd{0.805}{0.007} & \sd{0.769}{0.008} & \sd{0.754}{0.007} & \sd{0.813}{0.006} & \sd{0.041}{0.006} & \sd{0.073}{0.013} & \sd{0.048}{0.009} & \sd{0.638}{0.026} & \sd{0.638}{0.026} & \sd{0.640}{0.025} & \sd{0.641}{0.026} & \sd{0.035}{0.022} & \sd{0.041}{0.029} & \sd{0.048}{0.024} & \sd{0.779}{0.007} & \sd{0.683}{0.007} & \sd{0.683}{0.008} & \sd{0.686}{0.009} & \sd{0.045}{0.003} & \sd{0.028}{0.004} & \sd{0.048}{0.013} \\
\cmidrule(lr){2-23}
& \textbf{Average} & \sdb{0.842}{0.004} & \sdb{0.778}{0.005} & \sdb{0.800}{0.010} & \sdb{0.768}{0.003} & \sdb{0.060}{0.005} & \sdb{0.062}{0.002} & \sdb{0.042}{0.001}
& \sdb{0.640}{0.019} & \sdb{0.634}{0.020} & \sdb{0.637}{0.021} & \sdb{0.634}{0.020} & \sdb{0.116}{0.015} & \sdb{0.095}{0.010} & \sdb{0.105}{0.011}
& \sdb{0.808}{0.005} & \sdb{0.653}{0.009} & \sdb{0.743}{0.010} & \sdb{0.636}{0.007} & \sdb{0.035}{0.003} & \sdb{0.019}{0.007} & \sdb{0.036}{0.003} \\
\cmidrule(lr){1-23}
\multirow{10}{*}{LLM}
& Zero-Shot & \sd{0.767}{0.001} & \sd{0.728}{0.001} & \sd{0.717}{0.001} & \sd{0.765}{0.001} & \sd{0.213}{0.001} & \sd{0.161}{0.012} & \sd{0.159}{0.005} & \sd{0.643}{0.002} & \sd{0.642}{0.002} & \sd{0.661}{0.002} & \sd{0.655}{0.002} & \sd{0.608}{0.000} & \sd{0.778}{0.002} & \sd{0.623}{0.000} & \sd{0.779}{0.002} & \sd{0.554}{0.005} & \sd{0.643}{0.007} & \sd{0.555}{0.004} & \sd{0.018}{0.001} & \sd{0.023}{0.012} & \sd{0.014}{0.004} \\
\cmidrule(lr){2-23}
& Random    & \sd{0.789}{0.004} & \sd{0.729}{0.009} & \sd{0.722}{0.006} & \sd{0.741}{0.015} & \sd{0.124}{0.013} & \sd{0.076}{0.060} & \sd{0.075}{0.033} & \sd{0.630}{0.014} & \sd{0.616}{0.028} & \sd{0.627}{0.014} & \sd{0.619}{0.024} & \sd{0.393}{0.106} & \sd{0.524}{0.130} & \sd{0.405}{0.106} & \sd{0.674}{0.019} & \sd{0.610}{0.007} & \sd{0.613}{0.013} & \sd{0.653}{0.029} & \sd{0.100}{0.062} & \sd{0.087}{0.069} & \sd{0.083}{0.062} \\
& LF        & \sd{0.739}{0.005} & \sd{0.693}{0.012} & \sd{0.685}{0.013} & \sd{0.731}{0.020} & \sd{0.156}{0.034} & \sd{0.086}{0.036} & \sd{0.100}{0.034} & \sd{0.438}{0.082} & \sd{0.434}{0.082} & \sd{0.448}{0.092} & \sd{0.450}{0.086} & \sd{0.235}{0.193} & \sd{0.253}{0.154} & \sd{0.228}{0.177} & \sd{0.676}{0.026} & \sd{0.609}{0.020} & \sd{0.609}{0.015} & \sd{0.646}{0.015} & \sd{0.073}{0.039} & \sd{0.046}{0.042} & \sd{0.064}{0.037} \\
& S1        & \sd{0.764}{0.030} & \sd{0.713}{0.017} & \sd{0.706}{0.014} & \sd{0.741}{0.012} & \sd{0.151}{0.057} & \sd{0.091}{0.083} & \sd{0.098}{0.067} & \sd{0.608}{0.037} & \sd{0.606}{0.039} & \sd{0.616}{0.048} & \sd{0.613}{0.045} & \sd{0.551}{0.079} & \sd{0.675}{0.127} & \sd{0.562}{0.082} & \sd{0.679}{0.058} & \sd{0.608}{0.033} & \sd{0.612}{0.016} & \sd{0.643}{0.016} & \sd{0.067}{0.019} & \sd{0.070}{0.062} & \sd{0.056}{0.033} \\
& S2        & \sd{0.785}{0.025} & \sd{0.721}{0.005} & \sd{0.725}{0.022} & \sd{0.734}{0.032} & \sd{0.108}{0.020} & \sd{0.055}{0.025} & \sd{0.058}{0.014} & \sd{0.612}{0.032} & \sd{0.611}{0.031} & \sd{0.615}{0.026} & \sd{0.615}{0.027} & \sd{0.505}{0.057} & \sd{0.631}{0.048} & \sd{0.516}{0.058} & \sd{0.675}{0.018} & \sd{0.612}{0.008} & \sd{0.615}{0.014} & \sd{0.656}{0.030} & \sd{0.104}{0.060} & \sd{0.076}{0.075} & \sd{0.082}{0.063} \\
& S3        & \sd{0.814}{0.008} & \sd{0.733}{0.012} & \sd{0.756}{0.017} & \sd{0.721}{0.024} & \sd{0.076}{0.018} & \sd{0.045}{0.025} & \sd{0.037}{0.017} & \sd{0.615}{0.034} & \sd{0.608}{0.038} & \sd{0.626}{0.027} & \sd{0.619}{0.031} & \sd{0.553}{0.151} & \sd{0.684}{0.145} & \sd{0.565}{0.148} & \sd{0.676}{0.021} & \sd{0.611}{0.008} & \sd{0.614}{0.013} & \sd{0.654}{0.028} & \sd{0.104}{0.058} & \sd{0.091}{0.071} & \sd{0.086}{0.059} \\
& FCG-S1    & \sd{0.810}{0.015} & \sd{0.681}{0.088} & \sd{0.809}{0.058} & \sd{0.672}{0.085} & \sd{0.068}{0.053} & \sd{0.054}{0.049} & \sd{0.042}{0.041} & \sd{0.594}{0.034} & \sd{0.504}{0.093} & \sd{0.642}{0.020} & \sd{0.563}{0.043} & \sd{0.155}{0.149} & \sd{0.224}{0.198} & \sd{0.161}{0.155} & \sd{0.645}{0.052} & \sd{0.597}{0.035} & \sd{0.613}{0.012} & \sd{0.659}{0.012} & \sd{0.014}{0.010} & \sd{0.036}{0.028} & \sd{0.040}{0.022} \\
& FCG-S2    & \sd{0.814}{0.022} & \sd{0.736}{0.017} & \sd{0.761}{0.031} & \sd{0.727}{0.027} & \sd{0.112}{0.018} & \sd{0.106}{0.030} & \sd{0.082}{0.006} & \sd{0.621}{0.026} & \sd{0.619}{0.026} & \sd{0.638}{0.024} & \sd{0.633}{0.024} & \sd{0.609}{0.009} & \sd{0.755}{0.020} & \sd{0.622}{0.010} & \sd{0.653}{0.036} & \sd{0.595}{0.020} & \sd{0.604}{0.008} & \sd{0.644}{0.004} & \sd{0.006}{0.004} & \sd{0.029}{0.031} & \sd{0.033}{0.027} \\
& FCG-S3    & \sd{0.804}{0.017} & \sd{0.659}{0.064} & \sd{0.792}{0.050} & \sd{0.644}{0.059} & \sd{0.056}{0.032} & \sd{0.047}{0.016} & \sd{0.035}{0.019} & \sd{0.604}{0.036} & \sd{0.502}{0.091} & \sd{0.683}{0.022} & \sd{0.567}{0.042} & \sd{0.111}{0.079} & \sd{0.185}{0.126} & \sd{0.118}{0.083} & \sd{0.723}{0.029} & \sd{0.579}{0.035} & \sd{0.586}{0.029} & \sd{0.580}{0.035} & \sd{0.077}{0.036} & \sd{0.146}{0.080} & \sd{0.092}{0.047} \\
\cmidrule(lr){2-23}
& \textbf{Average} & \sdb{0.790}{0.012} & \sdb{0.708}{0.022} & \sdb{0.744}{0.009} & \sdb{0.714}{0.018} & \sdb{0.106}{0.006} & \sdb{0.070}{0.018} & \sdb{0.066}{0.010}
& \sdb{0.590}{0.014} & \sdb{0.563}{0.007} & \sdb{0.612}{0.028} & \sdb{0.585}{0.012} & \sdb{0.389}{0.041} & \sdb{0.491}{0.030} & \sdb{0.397}{0.038}
& \sdb{0.675}{0.006} & \sdb{0.603}{0.007} & \sdb{0.608}{0.010} & \sdb{0.642}{0.016} & \sdb{0.068}{0.031} & \sdb{0.073}{0.049} & \sdb{0.067}{0.039} \\
\bottomrule
\end{tabular}
}
\end{table*}

We analyze the results from both fairness and predictive performance.

\noindent\textbf{Fairness.}
Regarding fairness, the traditional ML paradigm consistently achieves lower bias levels than the LLM paradigm across all datasets. Under identical experimental settings across six tasks, traditional ML reduces SPD, EOD, and AOD by 48.3\%, 59.6\%, and 51.6\%, respectively, compared to the LLM paradigm. The gap is especially large on \textit{Compas-Race}, where ML reduces EOD by 80.7\% relative to the LLM average. Even when advanced demonstration strategies such as FCG-S3 improve fairness within the LLM paradigm, the remaining bias is still substantial compared with traditional ML methods. For example, on \textit{Compas-Race}, FCG-S3 reduces EOD by 76.2\% relative to the Zero-Shot baseline, but traditional ML still achieves a further 88.1\% reduction relative to FCG-S3. Consistent with these observations, ML is significantly better in 13 out of 18 fairness comparisons (72.2\%).

\noindent\textbf{Predictive performance.} Overall, the traditional ML paradigm consistently demonstrates superior predictive performance compared to the LLM paradigm across all six tasks. As shown in Table~\ref{tab:rq1_overall}, ML achieves higher average Accuracy, Precision, and F1-score in every configuration, while Recall is also higher on average overall. Across all six tasks, traditional ML achieves 11.1\% higher Accuracy, 9.3\% higher F1-score, 11.3\% higher Precision, and 4.6\% higher Recall than the LLM paradigm. This advantage is statistically significant in 20 out of 24 predictive-performance comparisons (83.3\%). Overall, these results confirm that traditional ML-based methods remain more effective than current LLM-based mitigation methods in both preserving predictive performance and reducing fairness disparities under practical tabular settings.

\finding{Under identical real-world experimental settings, the traditional ML paradigm consistently outperforms the LLM paradigm in both fairness and predictive performance. Across six tasks, traditional ML reduces SPD, EOD, and AOD by 48.3\%, 59.6\%, and 51.6\%, respectively, while achieving 11.1\% higher Accuracy, 9.3\% higher F1-score, 11.3\% higher Precision, and 4.6\% higher Recall on average. Moreover, ML is significantly better in 72.2\% of the fairness comparisons and 83.3\% of the predictive comparisons. Therefore, traditional ML remains the more effective and practical paradigm for tabular bias mitigation.}

\subsection{RQ2: Does the comparison depend on the choice of LLM?}
\subsubsection{Motivation and Methodology.} This RQ examines whether the overall comparison between the ML and LLM paradigms depends on the choice of LLM. To answer this question, we conduct an additional targeted validation using four advanced LLMs from major AI vendors: GPT-5~\cite{gpt5}, Gemini-2.5-flash~\cite{gemini2.5}, DeepSeek-v3.2~\cite{deepseek3.2}, and Qwen3-Max-Thinking~\cite{qwen3max}. These models cover both open-source and closed-source architectures from leading AI vendors (OpenAI, Google, DeepSeek, Alibaba), which are widely adopted in real-world applications~\cite{DBLP:journals/corr/abs-2303-18223, ChiangZ0ALLZ0JG24}. We focus this validation on \textit{Adult-Sex} and \textit{Adult-Race} because Adult is one of the most widely used and representative benchmark datasets in fairness research~\cite{liu2024confronting, hort2024bias}. Table~\ref{tab:rq2_consistency} reports the detailed results. For each LLM and each configuration, we average results over the eight mitigation methods to obtain the overall LLM results, and then compare them with the corresponding ML results from RQ1. We further assess statistical significance over the 56 displayed advanced-LLM comparisons ($56 = 2$ tasks $\times 4$ LLMs $\times 7$ metrics).

\begin{table*}[t]
\centering
\scriptsize
\caption{(RQ2) Predictive performance and fairness results of LLM-based mitigation methods across four advanced LLMs (QW: Qwen3-Max-Thinking, GPT: GPT-5, DS: DeepSeek-v3.2, and GM: Gemini-2.5-Flash) on \textit{Adult-Sex} and \textit{Adult-Race}. Even with stronger LLMs, the overall conclusion remains unchanged.}
\label{tab:rq2_consistency}
\setlength{\tabcolsep}{1.5pt}
\resizebox{\textwidth}{!}{
\renewcommand{\arraystretch}{0.95}
\begin{tabular}{l | llll | llll | llll | llll | llll | llll | llll}
\toprule
\multirow{2}{*}{\textbf{Method}} & \multicolumn{4}{c|}{\textbf{Acc.} $\uparrow$} & \multicolumn{4}{c|}{\textbf{F1} $\uparrow$} & \multicolumn{4}{c|}{\textbf{Prec.} $\uparrow$} & \multicolumn{4}{c|}{\textbf{Rec.} $\uparrow$} & \multicolumn{4}{c|}{\textbf{SPD} $\downarrow$} & \multicolumn{4}{c|}{\textbf{EOD} $\downarrow$} & \multicolumn{4}{c}{\textbf{AOD} $\downarrow$} \\
\cmidrule(lr){2-5} \cmidrule(lr){6-9} \cmidrule(lr){10-13} \cmidrule(lr){14-17} \cmidrule(lr){18-21} \cmidrule(lr){22-25} \cmidrule(lr){26-29}
& QW & GPT & DS & GM & QW & GPT & DS & GM & QW & GPT & DS & GM & QW & GPT & DS & GM & QW & GPT & DS & GM & QW & GPT & DS & GM & QW & GPT & DS & GM \\
\midrule
\multicolumn{29}{c}{Adult-Sex} \\
\midrule
Zero-Shot & 0.824 & 0.833 & 0.750 & 0.777 & 0.762 & 0.771 & 0.712 & 0.743 & 0.764 & 0.777 & 0.704 & 0.732 & 0.759 & 0.766 & 0.760 & 0.796 & 0.250 & 0.262 & 0.445 & 0.290 & 0.289 & 0.316 & 0.492 & 0.142 & 0.210 & 0.227 & 0.414 & 0.166 \\
\midrule
Random    & 0.803 & 0.829 & 0.811 & 0.841 & 0.743 & 0.756 & 0.744 & 0.779 & 0.738 & 0.780 & 0.747 & 0.791 & 0.751 & 0.743 & 0.743 & 0.770 & 0.193 & 0.204 & 0.168 & 0.207 & 0.129 & 0.227 & 0.096 & 0.185 & 0.111 & 0.160 & 0.084 & 0.138 \\
LF        & 0.618 & 0.391 & 0.465 & 0.646 & 0.602 & 0.386 & 0.463 & 0.610 & 0.651 & 0.510 & 0.586 & 0.636 & 0.696 & 0.502 & 0.596 & 0.666 & 0.325 & 0.178 & 0.171 & 0.302 & 0.246 & 0.129 & 0.132 & 0.277 & 0.254 & 0.154 & 0.137 & 0.261 \\
S1        & 0.780 & 0.805 & 0.787 & 0.836 & 0.733 & 0.759 & 0.737 & 0.783 & 0.721 & 0.745 & 0.725 & 0.780 & 0.763 & 0.785 & 0.761 & 0.786 & 0.206 & 0.251 & 0.205 & 0.220 & 0.116 & 0.157 & 0.095 & 0.166 & 0.111 & 0.151 & 0.102 & 0.135 \\
S2        & 0.758 & 0.775 & 0.775 & 0.829 & 0.722 & 0.743 & 0.727 & 0.782 & 0.715 & 0.734 & 0.717 & 0.771 & 0.774 & 0.801 & 0.757 & 0.799 & 0.271 & 0.345 & 0.198 & 0.271 & 0.160 & 0.249 & 0.090 & 0.198 & 0.167 & 0.243 & 0.097 & 0.176 \\
S3        & 0.722 & 0.720 & 0.755 & 0.790 & 0.692 & 0.693 & 0.709 & 0.754 & 0.699 & 0.702 & 0.699 & 0.742 & 0.759 & 0.767 & 0.745 & 0.801 & 0.179 & 0.046 & 0.078 & 0.222 & 0.053 & 0.079 & 0.058 & 0.100 & 0.069 & 0.081 & 0.042 & 0.107 \\
FCG-S1    & 0.818 & 0.833 & 0.801 & 0.836 & 0.746 & 0.732 & 0.727 & 0.777 & 0.758 & 0.814 & 0.749 & 0.782 & 0.740 & 0.708 & 0.729 & 0.775 & 0.125 & 0.140 & 0.125 & 0.220 & 0.069 & 0.106 & 0.016 & 0.166 & 0.052 & 0.079 & 0.030 & 0.138 \\
FCG-S2    & 0.772 & 0.818 & 0.818 & 0.828 & 0.734 & 0.776 & 0.760 & 0.782 & 0.723 & 0.764 & 0.756 & 0.770 & 0.780 & 0.805 & 0.765 & 0.801 & 0.335 & 0.332 & 0.182 & 0.289 & 0.252 & 0.266 & 0.086 & 0.230 & 0.243 & 0.240 & 0.084 & 0.200 \\
FCG-S3    & 0.795 & 0.827 & 0.803 & 0.828 & 0.746 & 0.767 & 0.731 & 0.779 & 0.737 & 0.769 & 0.737 & 0.770 & 0.772 & 0.766 & 0.729 & 0.792 & 0.218 & 0.163 & 0.119 & 0.239 & 0.147 & 0.074 & 0.089 & 0.167 & 0.130 & 0.066 & 0.062 & 0.145 \\
\midrule
\textbf{Average}   & 0.758 & 0.750 & 0.752 & 0.804 & 0.715 & 0.702 & 0.700 & 0.756 & 0.718 & 0.727 & 0.714 & 0.755 & 0.754 & 0.735 & 0.728 & 0.774 & 0.231 & 0.207 & 0.156 & 0.246 & 0.147 & 0.161 & 0.083 & 0.186 & 0.142 & 0.147 & 0.080 & 0.162 \\
\midrule
\textbf{Overall} & \multicolumn{4}{c|}{\textbf{0.766}} & \multicolumn{4}{c|}{\textbf{0.718}} & \multicolumn{4}{c|}{\textbf{0.729}} & \multicolumn{4}{c|}{\textbf{0.748}} & \multicolumn{4}{c|}{\textbf{0.210}} & \multicolumn{4}{c|}{\textbf{0.144}} & \multicolumn{4}{c}{\textbf{0.133}} \\
\midrule
\multicolumn{29}{c}{Adult-Race} \\
\midrule
Zero-Shot & 0.824 & 0.834 & 0.746 & 0.777 & 0.761 & 0.772 & 0.708 & 0.743 & 0.765 & 0.779 & 0.701 & 0.732 & 0.758 & 0.767 & 0.757 & 0.797 & 0.149 & 0.163 & 0.305 & 0.193 & 0.165 & 0.219 & 0.336 & 0.098 & 0.124 & 0.153 & 0.287 & 0.119 \\
\midrule
Random    & 0.819 & 0.827 & 0.813 & 0.842 & 0.749 & 0.753 & 0.747 & 0.780 & 0.760 & 0.777 & 0.752 & 0.792 & 0.741 & 0.741 & 0.745 & 0.771 & 0.130 & 0.141 & 0.122 & 0.125 & 0.139 & 0.189 & 0.096 & 0.117 & 0.104 & 0.130 & 0.082 & 0.089 \\
LF        & 0.707 & 0.536 & 0.569 & 0.790 & 0.674 & 0.523 & 0.558 & 0.746 & 0.679 & 0.600 & 0.623 & 0.736 & 0.735 & 0.624 & 0.659 & 0.780 & 0.207 & 0.167 & 0.128 & 0.167 & 0.142 & 0.100 & 0.040 & 0.100 & 0.153 & 0.139 & 0.074 & 0.106 \\
S1        & 0.745 & 0.797 & 0.757 & 0.831 & 0.707 & 0.754 & 0.715 & 0.778 & 0.710 & 0.747 & 0.714 & 0.774 & 0.757 & 0.785 & 0.758 & 0.784 & 0.163 & 0.174 & 0.163 & 0.140 & 0.069 & 0.107 & 0.067 & 0.111 & 0.094 & 0.113 & 0.094 & 0.094 \\
S2        & 0.705 & 0.778 & 0.770 & 0.828 & 0.681 & 0.741 & 0.729 & 0.780 & 0.703 & 0.741 & 0.726 & 0.771 & 0.763 & 0.789 & 0.771 & 0.796 & 0.228 & 0.172 & 0.152 & 0.147 & 0.086 & 0.094 & 0.053 & 0.114 & 0.140 & 0.106 & 0.080 & 0.098 \\
S3        & 0.725 & 0.796 & 0.784 & 0.830 & 0.700 & 0.757 & 0.738 & 0.784 & 0.710 & 0.746 & 0.728 & 0.773 & 0.779 & 0.798 & 0.768 & 0.802 & 0.185 & 0.134 & 0.120 & 0.139 & 0.054 & 0.060 & 0.028 & 0.075 & 0.099 & 0.071 & 0.050 & 0.076 \\
FCG-S1    & 0.800 & 0.821 & 0.810 & 0.841 & 0.750 & 0.760 & 0.744 & 0.779 & 0.741 & 0.776 & 0.751 & 0.789 & 0.771 & 0.766 & 0.747 & 0.770 & 0.141 & 0.135 & 0.116 & 0.131 & 0.088 & 0.111 & 0.087 & 0.141 & 0.087 & 0.093 & 0.075 & 0.102 \\
FCG-S2    & 0.768 & 0.828 & 0.812 & 0.830 & 0.728 & 0.767 & 0.747 & 0.780 & 0.725 & 0.779 & 0.753 & 0.772 & 0.774 & 0.769 & 0.751 & 0.793 & 0.255 & 0.159 & 0.133 & 0.151 & 0.242 & 0.182 & 0.094 & 0.112 & 0.217 & 0.135 & 0.087 & 0.100 \\
FCG-S3    & 0.800 & 0.829 & 0.786 & 0.834 & 0.751 & 0.767 & 0.725 & 0.779 & 0.742 & 0.771 & 0.717 & 0.777 & 0.775 & 0.765 & 0.737 & 0.782 & 0.166 & 0.101 & 0.091 & 0.135 & 0.121 & 0.072 & 0.041 & 0.112 & 0.114 & 0.061 & 0.042 & 0.091 \\
\midrule
\textbf{Average} & 0.759 & 0.776 & 0.763 & 0.828 & 0.718 & 0.728 & 0.713 & 0.776 & 0.721 & 0.742 & 0.720 & 0.773 & 0.762 & 0.755 & 0.742 & 0.785 & 0.184 & 0.148 & 0.128 & 0.142 & 0.118 & 0.114 & 0.063 & 0.110 & 0.126 & 0.106 & 0.073 & 0.095 \\
\midrule
\textbf{Overall} & \multicolumn{4}{c|}{\textbf{0.782}} & \multicolumn{4}{c|}{\textbf{0.733}} & \multicolumn{4}{c|}{\textbf{0.739}} & \multicolumn{4}{c|}{\textbf{0.761}} & \multicolumn{4}{c|}{\textbf{0.151}} & \multicolumn{4}{c|}{\textbf{0.101}} & \multicolumn{4}{c}{\textbf{0.100}} \\
\bottomrule
\end{tabular}
}
\end{table*}

\subsubsection{Results.} The overall conclusion remains consistent with RQ1. Across both \textit{Adult-Sex} and \textit{Adult-Race}, traditional ML still outperforms the overall LLM averages on all three fairness metrics and all four predictive metrics. Specifically, ML reduces SPD, EOD, and AOD by 48.6\%, 35.4\%, and 48.9\% on \textit{Adult-Sex}, and by 60.3\%, 38.6\%, and 58.0\% on \textit{Adult-Race}, respectively. It also improves Accuracy, F1-score, Precision, and Recall by 8.9\%, 6.4\%, 9.1\%, and 0.5\% on \textit{Adult-Sex}, and by 7.7\%, 6.1\%, 8.3\%, and 0.9\% on \textit{Adult-Race}, respectively. This advantage is statistically significant in 46 of the 56 comparisons (82.1\%). Therefore, even after replacing GPT-4o-mini with stronger models, the overall conclusion of RQ1 remains unchanged.

Another notable observation is that even the largest reasoning-oriented model in our evaluation does not show clear mitigation advantages over traditional ML. Although Qwen3-Max-Thinking is the largest model we study, it still lags behind ML. On \textit{Adult-Sex}, ML reduces SPD, EOD, and AOD by 48.3\%, 35.5\%, and 47.7\%, respectively, while achieving 8.9\%, 6.0\%, 7.8\%, and 0.4\% higher Accuracy, F1-score, Precision, and Recall. A similar pattern also appears on \textit{Adult-Race}. This suggests that simply scaling up LLM size or reasoning capability is insufficient to address the challenges of tabular bias mitigation.

\finding{Even when stronger advanced LLMs are considered, the overall conclusion remains unchanged. Traditional ML is significantly better in 82.1\% of the comparisons, and the advanced LLMs do not overturn ML's overall advantage in either fairness or predictive performance. This shows that the advantage of the ML paradigm remains consistent across different advanced LLMs.}

\subsection{RQ3: How do evaluation settings affect the effectiveness of LLM-based methods?}
\subsubsection{Motivation and Methodology.} RQ1 and RQ2 show that LLM-based mitigation methods underperform traditional ML-based methods under our evaluation setting, and that this conclusion remains the same across different LLMs. 
However, existing LLM-based bias mitigation studies~\cite{wang2023decodingtrust, hu2024strategic, liu2024confronting} often report promising results. Furthermore, we find that these studies~\cite{wang2023decodingtrust, hu2024strategic, liu2024confronting} typically conduct evaluations on artificially balanced test sets, where samples are evenly distributed across demographic groups and labels. In contrast, our evaluation follows real-world data distributions, which are inherently imbalanced. This observation leads us to hypothesize that evaluation settings, particularly the use of balanced versus imbalanced test sets, may substantially influence the observed fairness performance. Therefore, in this RQ, we systematically investigate the impact of evaluation settings on the effectiveness of LLM-based mitigation methods.

To answer this question, we focus on \textit{Adult-Sex} and \textit{Adult-Race} and compare the same LLM-based mitigation methods under two test-set settings. The \textbf{balanced-test setting} follows previous LLM evaluation settings and samples 512 test instances with balanced demographic and label proportions. The \textbf{random-test setting} samples 512 test instances uniformly from the Adult test set and thus preserves the natural distribution. Both test-set settings are repeated three times with different random seeds. To assess whether this overall pattern is statistically significant, we aggregate the raw results across the nine LLM-based methods and three repeated runs, giving 27 raw values for each fairness metric under each task. A difference is considered statistically significant when the resulting $p$-value is below 0.05, and the balanced-test setting is considered better when it yields a lower mean fairness value.

\begin{table*}[t]
\centering
\scriptsize
\renewcommand{\arraystretch}{0.95}
\caption{(RQ3) Comparison of LLM-based mitigation methods under the balanced-test setting and the random-test setting on Adult. The \textit{balanced-test setting} uses balanced demographic and label proportions, while the \textit{random-test setting} preserves the original test distribution. Each value is averaged over three runs.}
\label{tab:rq3_balanced_vs_random}
\setlength{\tabcolsep}{1pt}
\resizebox{\textwidth}{!}{
\begin{tabular}{l|lllllll|lllllll|lllllll|lllllll}
\toprule
\multirow{3}{*}{Method}
& \multicolumn{14}{c|}{Adult-Sex}
& \multicolumn{14}{c}{Adult-Race} \\
\cmidrule(lr){2-15}\cmidrule(lr){16-29}
& \multicolumn{7}{c|}{Balanced}
& \multicolumn{7}{c|}{Random}
& \multicolumn{7}{c|}{Balanced}
& \multicolumn{7}{c}{Random} \\
\cmidrule(lr){2-8}\cmidrule(lr){9-15}\cmidrule(lr){16-22}\cmidrule(lr){23-29}
& Acc.$\uparrow$ & F1$\uparrow$ & Prec.$\uparrow$ & Rec.$\uparrow$ & SPD$\downarrow$ & EOD$\downarrow$ & AOD$\downarrow$
& Acc.$\uparrow$ & F1$\uparrow$ & Prec.$\uparrow$ & Rec.$\uparrow$ & SPD$\downarrow$ & EOD$\downarrow$ & AOD$\downarrow$
& Acc.$\uparrow$ & F1$\uparrow$ & Prec.$\uparrow$ & Rec.$\uparrow$ & SPD$\downarrow$ & EOD$\downarrow$ & AOD$\downarrow$
& Acc.$\uparrow$ & F1$\uparrow$ & Prec.$\uparrow$ & Rec.$\uparrow$ & SPD$\downarrow$ & EOD$\downarrow$ & AOD$\downarrow$ \\
\midrule
Zero-Shot  & 0.737 & 0.735 & 0.743 & 0.737 & 0.143 & 0.148 & 0.143
& 0.785 & 0.732 & 0.720 & 0.754 & 0.258 & 0.158 & 0.159
& 0.760 & 0.758 & 0.771 & 0.760 & 0.175 & 0.203 & 0.175
& 0.785 & 0.732 & 0.719 & 0.755 & 0.192 & 0.256 & 0.186 \\
Random     & 0.745 & 0.741 & 0.761 & 0.745 & 0.081 & 0.112 & 0.083
& 0.793 & 0.727 & 0.722 & 0.733 & 0.152 & 0.042 & 0.054
& 0.763 & 0.762 & 0.769 & 0.763 & 0.037 & 0.076 & 0.068
& 0.788 & 0.731 & 0.721 & 0.748 & 0.121 & 0.220 & 0.147 \\
LF         & 0.727 & 0.726 & 0.730 & 0.727 & 0.009 & 0.029 & 0.020
& 0.741 & 0.685 & 0.675 & 0.711 & 0.102 & 0.028 & 0.025
& 0.754 & 0.753 & 0.758 & 0.754 & 0.039 & 0.065 & 0.057
& 0.783 & 0.727 & 0.715 & 0.746 & 0.095 & 0.162 & 0.106 \\
S1      & 0.757 & 0.757 & 0.757 & 0.757 & 0.035 & 0.039 & 0.043
& 0.749 & 0.702 & 0.692 & 0.738 & 0.132 & 0.060 & 0.054
& 0.740 & 0.739 & 0.745 & 0.740 & 0.043 & 0.081 & 0.074
& 0.776 & 0.721 & 0.710 & 0.742 & 0.104 & 0.067 & 0.067 \\
S2      & 0.746 & 0.746 & 0.746 & 0.746 & 0.068 & 0.065 & 0.068
& 0.753 & 0.708 & 0.699 & 0.749 & 0.169 & 0.086 & 0.084
& 0.760 & 0.760 & 0.761 & 0.760 & 0.030 & 0.055 & 0.056
& 0.769 & 0.722 & 0.710 & 0.756 & 0.111 & 0.159 & 0.105 \\
S3      & 0.744 & 0.742 & 0.751 & 0.744 & 0.043 & 0.047 & 0.059
& 0.796 & 0.735 & 0.727 & 0.747 & 0.165 & 0.031 & 0.054
& 0.729 & 0.723 & 0.752 & 0.729 & 0.060 & 0.107 & 0.065
& 0.803 & 0.733 & 0.734 & 0.731 & 0.114 & 0.201 & 0.119 \\
FCG-S1  & 0.673 & 0.637 & 0.772 & 0.673 & 0.042 & 0.070 & 0.047
& 0.829 & 0.728 & 0.804 & 0.703 & 0.120 & 0.161 & 0.101
& 0.697 & 0.665 & 0.769 & 0.697 & 0.048 & 0.076 & 0.048
& 0.813 & 0.744 & 0.747 & 0.740 & 0.133 & 0.208 & 0.132 \\
FCG-S2  & 0.741 & 0.733 & 0.771 & 0.741 & 0.159 & 0.206 & 0.159
& 0.810 & 0.745 & 0.750 & 0.747 & 0.215 & 0.151 & 0.132
& 0.739 & 0.735 & 0.755 & 0.739 & 0.046 & 0.078 & 0.074
& 0.784 & 0.721 & 0.713 & 0.733 & 0.137 & 0.198 & 0.133 \\
FCG-S3  & 0.707 & 0.685 & 0.776 & 0.707 & 0.083 & 0.117 & 0.083
& 0.842 & 0.749 & 0.825 & 0.721 & 0.144 & 0.080 & 0.066
& 0.627 & 0.576 & 0.744 & 0.627 & 0.035 & 0.070 & 0.043
& 0.797 & 0.644 & 0.761 & 0.625 & 0.029 & 0.080 & 0.049 \\
\midrule
\textbf{Average} & \textbf{0.731} & \textbf{0.722} & \textbf{0.757} & \textbf{0.731} & \textbf{0.074} & \textbf{0.093} & \textbf{0.078}
& \textbf{0.789} & \textbf{0.723} & \textbf{0.735} & \textbf{0.734} & \textbf{0.162} & \textbf{0.089} & \textbf{0.081}
& \textbf{0.730} & \textbf{0.719} & \textbf{0.758} & \textbf{0.730} & \textbf{0.057} & \textbf{0.090} & \textbf{0.073}
& \textbf{0.789} & \textbf{0.719} & \textbf{0.725} & \textbf{0.731} & \textbf{0.115} & \textbf{0.172} & \textbf{0.116} \\
\bottomrule
\end{tabular}
}
\end{table*}

\subsubsection{Results.} Table~\ref{tab:rq3_balanced_vs_random} shows that the balanced-test setting usually reports better fairness than the random-test setting. Specifically, 77.8\% of the method-level fairness comparisons numerically favor the balanced-test setting. The balanced-test setting is significantly better in 66.7\% of the fairness comparisons. Averaged over the eight mitigation methods, SPD, EOD, and AOD increase by 138.4\%, 49.7\%, and 36.2\%, respectively, when evaluation shifts from the balanced-test setting to the random-test setting.

The balanced-test setting removes much of the demographic skew in the original data, which makes fairness disparities appear smaller than under the natural distribution. Once evaluation returns to the random-test setting, this apparent advantage weakens substantially. This helps explain why prior LLM studies can report more favorable fairness outcomes under balanced evaluation, while our results in RQ1 and RQ2 remain much less favorable under more realistic test distributions.

\finding{Fairness in the LLM paradigm is highly sensitive to the test setting. Across Adult-Sex and Adult-Race, 77.8\% of the method-level fairness comparisons numerically favor the balanced-test setting, and 66.7\% of the fairness comparisons show statistically significant advantages for the balanced-test setting. Under the random-test setting, SPD, EOD, and AOD increase by 138.4\%, 49.7\%, and 36.2\%, respectively. This suggests that the favorable fairness reported in previous LLM studies may rely on balanced evaluation and may not generalize to more realistic test distributions.}

\subsection{RQ4: Can fine-tuning on the full training data improve LLM-based methods?}
\subsubsection{Motivation and Methodology.} 
Beyond the influence of test data, we further investigate whether the performance gap between LLM- and ML-based methods can be attributed to differences in training data utilization. Existing LLM-based bias mitigation methods~\cite{hu2024strategic, wang2023decodingtrust} primarily rely on in-context learning, which leverages only a small subset of training data as demonstrations, whereas ML-based methods are trained on the full dataset. This discrepancy raises the question of whether limited access to training data constrains the effectiveness of LLM-based approaches. To examine this, we move beyond in-context learning and instead adopt supervised fine-tuning, which allows LLMs to fully leverage the entire training dataset. Specifically, we fine-tune LLMs on the full training set and evaluate whether such broader data exposure can improve their effectiveness.

Based on this motivation, we study both regular fine-tuning and fine-tuning combined with traditional data-level pre-processing, and examine whether these gains are enough to challenge the traditional ML paradigm.

\noindent \textbf{Fine-tuning-based strategies.} We evaluate four fine-tuning-based strategies using supervised LLM fine-tuning. Unlike in-context learning, which relies on limited demonstrations without updating model parameters, fine-tuning adapts the LLM using the full training dataset.

\begin{itemize}[leftmargin=*]
\item \textbf{Regular Fine-tuning.} The LLM is fine-tuned in a standard supervised manner using the original tabular training data without any fairness intervention.
\item \textbf{Pre-processed Fine-tuning.} Traditional pre-processing bias mitigation techniques are first applied to the training data, after which the LLM is fine-tuned on the modified dataset. Specifically, we evaluate fine-tuning in combination with COT, LTDD, and FairMask to examine whether data-level debiasing can influence fairness outcomes in the fine-tuned LLM.
\end{itemize}

We evaluate these strategies on all six tasks using two lightweight and economical LLMs, GPT-4o-mini~\cite{gpt4omini} and Qwen2.5-7B~\cite{qwen2.5}. For each setting, we keep the train-test split fixed for fine-tuning and repeat evaluation three times on the same test set. Table~\ref{tab:rq4_cross_paradigm} reports the detailed results. To assess statistical significance, we compare the raw outcomes of fine-tuning with the in-context learning results in RQ1 and with the traditional ML results in RQ1. Across all six tasks, this gives 24 predictive-performance comparisons and 18 fairness comparisons.

\begin{table*}[t]
\centering
\scriptsize
\renewcommand{\arraystretch}{0.95}
\caption{(RQ4) Fine-tuning-based mitigation results. \textit{Regular} denotes standard supervised fine-tuning on the original training data. \textit{COT}, \textit{LTDD}, and \textit{FairMask} denote pre-processed fine-tuning, where the corresponding pre-processing method is applied before fine-tuning. Each value is averaged over three repeated tests.}
\label{tab:rq4_cross_paradigm}
\setlength{\tabcolsep}{2pt}
\resizebox{\textwidth}{!}{
\begin{tabular}{l|l|lllllll|lllllll|lllllll}
\toprule
\multirow{2}{*}{Model} & \multirow{2}{*}{Method}
& \multicolumn{7}{c|}{Adult-Sex}
& \multicolumn{7}{c|}{Compas-Sex}
& \multicolumn{7}{c}{Credit-Sex} \\
\cmidrule(lr){3-9}\cmidrule(lr){10-16}\cmidrule(lr){17-23}
& & Acc.$\uparrow$ & F1$\uparrow$ & Prec.$\uparrow$ & Rec.$\uparrow$ & SPD$\downarrow$ & EOD$\downarrow$ & AOD$\downarrow$
  & Acc.$\uparrow$ & F1$\uparrow$ & Prec.$\uparrow$ & Rec.$\uparrow$ & SPD$\downarrow$ & EOD$\downarrow$ & AOD$\downarrow$
  & Acc.$\uparrow$ & F1$\uparrow$ & Prec.$\uparrow$ & Rec.$\uparrow$ & SPD$\downarrow$ & EOD$\downarrow$ & AOD$\downarrow$ \\
\midrule
\multirow{4}{*}{GPT-4o-mini} & Regular & 0.871 & 0.814 & 0.844 & 0.794 & 0.168 & 0.055 & 0.055 & 0.675 & 0.671 & 0.671 & 0.670 & 0.224 & 0.199 & 0.183 & 0.821 & 0.674 & 0.766 & 0.648 & 0.013 & 0.006 & 0.017 \\
\cmidrule(lr){2-23}
 & COT & 0.871 & 0.815 & 0.841 & 0.796 & 0.161 & 0.031 & 0.041 & 0.672 & 0.656 & 0.675 & 0.658 & 0.043 & 0.010 & 0.014 & 0.812 & 0.652 & 0.747 & 0.630 & 0.008 & 0.034 & 0.020 \\
 & LTDD & 0.787 & 0.568 & 0.856 & 0.574 & 0.017 & 0.069 & 0.035 & 0.666 & 0.647 & 0.671 & 0.650 & 0.090 & 0.048 & 0.050 & 0.814 & 0.663 & 0.746 & 0.640 & 0.008 & 0.031 & 0.021 \\
 & FairMask & 0.867 & 0.803 & 0.847 & 0.777 & 0.133 & 0.016 & 0.024 & 0.666 & 0.653 & 0.666 & 0.653 & 0.148 & 0.132 & 0.112 & 0.821 & 0.679 & 0.759 & 0.654 & 0.016 & 0.024 & 0.016 \\
\cmidrule(lr){2-23}
 & Average & 0.849 & 0.750 & 0.847 & 0.735 & 0.120 & 0.043 & 0.039 & 0.670 & 0.657 & 0.671 & 0.658 & 0.126 & 0.097 & 0.090 & 0.817 & 0.667 & 0.755 & 0.643 & 0.011 & 0.024 & 0.018 \\
\cmidrule(lr){1-23}
\multirow{4}{*}{Qwen2.5-7B} & Regular & 0.871 & 0.821 & 0.834 & 0.811 & 0.203 & 0.119 & 0.099 & 0.644 & 0.591 & 0.686 & 0.615 & 0.134 & 0.060 & 0.119 & 0.808 & 0.601 & 0.782 & 0.590 & 0.016 & 0.008 & 0.018 \\
\cmidrule(lr){2-23}
 & COT & 0.872 & 0.817 & 0.841 & 0.800 & 0.176 & 0.067 & 0.064 & 0.650 & 0.603 & 0.686 & 0.622 & 0.085 & 0.035 & 0.059 & 0.818 & 0.653 & 0.770 & 0.629 & 0.006 & 0.002 & 0.014 \\
 & LTDD & 0.799 & 0.634 & 0.793 & 0.617 & 0.054 & 0.022 & 0.016 & 0.614 & 0.581 & 0.617 & 0.592 & 0.054 & 0.058 & 0.042 & 0.790 & 0.500 & 0.791 & 0.530 & 0.003 & 0.002 & 0.007 \\
 & FairMask & 0.870 & 0.815 & 0.840 & 0.797 & 0.182 & 0.083 & 0.074 & 0.619 & 0.535 & 0.690 & 0.583 & 0.061 & 0.030 & 0.041 & 0.817 & 0.648 & 0.774 & 0.625 & 0.009 & 0.004 & 0.009 \\
\cmidrule(lr){2-23}
 & Average & 0.853 & 0.772 & 0.827 & 0.756 & 0.154 & 0.073 & 0.063 & 0.632 & 0.577 & 0.670 & 0.603 & 0.084 & 0.046 & 0.065 & 0.808 & 0.601 & 0.779 & 0.594 & 0.009 & 0.004 & 0.012 \\
\midrule
\multicolumn{2}{c|}{\textbf{Overall}} & \textbf{0.851} & \textbf{0.761} & \textbf{0.837} & \textbf{0.746} & \textbf{0.137} & \textbf{0.058} & \textbf{0.051} & \textbf{0.651} & \textbf{0.617} & \textbf{0.670} & \textbf{0.630} & \textbf{0.105} & \textbf{0.071} & \textbf{0.078} & \textbf{0.813} & \textbf{0.634} & \textbf{0.767} & \textbf{0.618} & \textbf{0.010} & \textbf{0.014} & \textbf{0.015} \\
\midrule
\multirow{2}{*}{Model} & \multirow{2}{*}{Method}
& \multicolumn{7}{c|}{Adult-Race}
& \multicolumn{7}{c|}{Compas-Race}
& \multicolumn{7}{c}{Credit-Age} \\
\cmidrule(lr){3-9}\cmidrule(lr){10-16}\cmidrule(lr){17-23}
& & Acc.$\uparrow$ & F1$\uparrow$ & Prec.$\uparrow$ & Rec.$\uparrow$ & SPD$\downarrow$ & EOD$\downarrow$ & AOD$\downarrow$
  & Acc.$\uparrow$ & F1$\uparrow$ & Prec.$\uparrow$ & Rec.$\uparrow$ & SPD$\downarrow$ & EOD$\downarrow$ & AOD$\downarrow$
  & Acc.$\uparrow$ & F1$\uparrow$ & Prec.$\uparrow$ & Rec.$\uparrow$ & SPD$\downarrow$ & EOD$\downarrow$ & AOD$\downarrow$ \\
\midrule
\multirow{4}{*}{GPT-4o-mini} & Regular & 0.871 & 0.814 & 0.844 & 0.794 & 0.093 & 0.059 & 0.043 & 0.675 & 0.670 & 0.671 & 0.670 & 0.432 & 0.618 & 0.448 & 0.821 & 0.673 & 0.766 & 0.647 & 0.057 & 0.025 & 0.051 \\
\cmidrule(lr){2-23}
 & COT & 0.870 & 0.814 & 0.840 & 0.796 & 0.093 & 0.052 & 0.040 & 0.673 & 0.664 & 0.671 & 0.663 & 0.117 & 0.051 & 0.079 & 0.821 & 0.666 & 0.771 & 0.640 & 0.053 & 0.023 & 0.048 \\
 & LTDD & 0.834 & 0.754 & 0.792 & 0.732 & 0.058 & 0.021 & 0.019 & 0.572 & 0.554 & 0.631 & 0.598 & 0.087 & 0.152 & 0.110 & 0.804 & 0.586 & 0.775 & 0.579 & 0.043 & 0.014 & 0.085 \\
 & FairMask & 0.857 & 0.767 & 0.872 & 0.730 & 0.052 & 0.008 & 0.007 & 0.674 & 0.667 & 0.671 & 0.667 & 0.209 & 0.156 & 0.170 & 0.822 & 0.674 & 0.768 & 0.648 & 0.059 & 0.026 & 0.055 \\
\cmidrule(lr){2-23}
 & Average & 0.858 & 0.787 & 0.837 & 0.763 & 0.074 & 0.035 & 0.028 & 0.649 & 0.639 & 0.661 & 0.650 & 0.211 & 0.244 & 0.202 & 0.817 & 0.650 & 0.770 & 0.629 & 0.053 & 0.022 & 0.060 \\
\cmidrule(lr){1-23}
\multirow{4}{*}{Qwen2.5-7B} & Regular & 0.871 & 0.821 & 0.834 & 0.811 & 0.122 & 0.126 & 0.086 & 0.645 & 0.591 & 0.688 & 0.615 & 0.229 & 0.117 & 0.212 & 0.808 & 0.601 & 0.780 & 0.590 & 0.022 & 0.007 & 0.017 \\
\cmidrule(lr){2-23}
 & COT & 0.872 & 0.817 & 0.841 & 0.800 & 0.104 & 0.074 & 0.055 & 0.648 & 0.600 & 0.682 & 0.620 & 0.206 & 0.115 & 0.185 & 0.818 & 0.654 & 0.772 & 0.630 & 0.051 & 0.024 & 0.045 \\
 & LTDD & 0.756 & 0.607 & 0.653 & 0.597 & 0.043 & 0.046 & 0.033 & 0.620 & 0.618 & 0.618 & 0.619 & 0.092 & 0.137 & 0.123 & 0.781 & 0.450 & 0.748 & 0.505 & 0.004 & 0.001 & 0.008 \\
 & FairMask & 0.859 & 0.774 & 0.868 & 0.738 & 0.066 & 0.040 & 0.026 & 0.623 & 0.544 & 0.689 & 0.588 & 0.106 & 0.053 & 0.088 & 0.788 & 0.496 & 0.775 & 0.528 & 0.013 & 0.007 & 0.017 \\
\cmidrule(lr){2-23}
 & Average & 0.840 & 0.755 & 0.799 & 0.736 & 0.084 & 0.072 & 0.050 & 0.634 & 0.588 & 0.669 & 0.610 & 0.158 & 0.106 & 0.152 & 0.799 & 0.550 & 0.769 & 0.563 & 0.023 & 0.010 & 0.022 \\
\midrule
\multicolumn{2}{c|}{\textbf{Overall}} & \textbf{0.849} & \textbf{0.771} & \textbf{0.818} & \textbf{0.750} & \textbf{0.079} & \textbf{0.053} & \textbf{0.039} & \textbf{0.641} & \textbf{0.613} & \textbf{0.665} & \textbf{0.630} & \textbf{0.185} & \textbf{0.175} & \textbf{0.177} & \textbf{0.808} & \textbf{0.600} & \textbf{0.769} & \textbf{0.596} & \textbf{0.038} & \textbf{0.016} & \textbf{0.041} \\
\bottomrule
\end{tabular}
}
\end{table*}

\subsubsection{Results.}
\noindent\textbf{Compared with in-context learning.} Compared with the in-context learning results in RQ1, fine-tuning is more effective. Averaged over the pre-processed variants and the two lightweight models across all six tasks, SPD, EOD, and AOD decrease by 42.3\%, 54.0\%, and 50.4\%, respectively. Fine-tuning is significantly better in 72.2\% of fairness comparisons against the in-context learning results in RQ1. Meanwhile, Accuracy, F1-score, Precision, and Recall improve by 11.2\%, 4.4\%, 15.0\%, and 0.8\%, respectively, and fine-tuning is significantly better in 79.2\% of predictive-performance comparisons. These results indicate that giving the model access to the full training set is much more effective than relying on demonstrations alone.

\noindent\textbf{Effect of pre-processing.} Pre-processing further improves fine-tuning. Compared with regular fine-tuning, the pre-processed variants reduce SPD, EOD, and AOD by 40.9\%, 40.3\%, and 41.0\% on average for GPT-4o-mini, and by 39.7\%, 39.0\%, and 45.1\% for Qwen2.5-7B. These improvements come with only modest drops in predictive performance, with Accuracy and F1-score decreasing by 2.1\% and 5.2\% for GPT-4o-mini, and by 2.4\% and 6.1\% for Qwen2.5-7B.

\noindent\textbf{Compared with traditional ML.} Relative to traditional ML, the gains from fine-tuning remain limited. Compared with the ML results in RQ1, the overall fine-tuning results achieve better mean values in 50.0\% of predictive-performance comparisons and 61.1\% of fairness comparisons. However, fine-tuning is significantly better in only 37.5\% of predictive-performance comparisons and in none of the fairness comparisons. In addition, fine-tuning introduces extra monetary cost (on average \$8.32 per task and per strategy) and training time (on average 1.10 hours per task and per strategy), which further limits its practical advantage over traditional ML in tabular bias mitigation. These results suggest that although fine-tuning strengthens the LLM paradigm, its advantage over traditional ML remains limited in both effectiveness and practical cost.

\finding{Fine-tuning-based methods improve over in-context learning, and pre-processing further helps. However, gains over traditional ML remain limited. The overall fine-tuning results achieve better mean values in 61.1\% of fairness comparisons and 50.0\% of predictive-performance comparisons, but significant advantages appear in none of the fairness comparisons and in only 37.5\% of predictive-performance comparisons. Fine-tuning also introduces extra monetary cost and training time. These results suggest that although fine-tuning strengthens the LLM paradigm, its advantage over traditional ML remains limited in both effectiveness and practical cost.}

\section{Implications}
This section derives implications for future research and practice, based on our findings.

\noindent\textbf{(1) LLMs are not a silver bullet.} Although LLMs are currently receiving substantial attention, our findings (RQ1 and RQ2) show that, in tabular bias mitigation, traditional ML-based methods achieve stronger fairness and predictive performance than LLM-based methods. RQ4 further shows that, although fine-tuning on the full training data improves LLM-based methods, it still does not consistently outperform traditional ML-based mitigation. These suggest that LLMs should not be assumed to be a universal replacement for traditional bias mitigation techniques. For software engineers, this implies that adopting LLM-based methods should not be treated as a default upgrade in fairness-critical tabular applications; instead, method selection should remain evidence-driven and task-specific. More broadly, our findings highlight a general lesson for SE: the adoption of emerging paradigms such as LLMs should be guided by empirical evidence rather than prevailing trends. Even highly capable foundation models may not uniformly outperform well-established techniques in domain-specific tasks, underscoring the importance of rigorous evaluation when integrating new paradigms into software systems. 

\noindent\textbf{(2) Real-world adoption requires evidence from realistic evaluation settings.} Our findings from RQ3 show that the effectiveness of LLM-based bias mitigation methods is highly sensitive to evaluation settings. In particular, artificially balanced test distributions can substantially inflate fairness improvements, compared to evaluations under realistic, imbalanced data distributions, and may therefore create a misleading impression of real-world effectiveness. This has an important implication for both research and practice. For practitioners, fairness improvements demonstrated only under balanced test settings may not translate to real-world deployments, where class and demographic distributions are inherently skewed. Relying on such results may therefore lead to suboptimal or even misleading deployment decisions. For researchers, these results highlight that evaluation settings should be treated as an integral part of fairness claims. Methods intended for real-world use should be evaluated primarily under realistic data distributions, and results obtained under controlled or balanced settings should be interpreted with caution. More broadly, this finding underscores a general lesson for SE: the validity of empirical conclusions depends not only on the methods being evaluated, but also on how closely the evaluation setting reflects real-world operating conditions.

\noindent\textbf{(3) Cross-paradigm evaluation is essential.}  
Our study demonstrates that cross-paradigm comparisons are necessary to obtain a complete understanding of method effectiveness. Existing research on bias mitigation typically evaluates methods within a single paradigm (e.g., ML-based or LLM-based), which may lead to an incomplete or overly optimistic view of their effectiveness. By directly comparing ML- and LLM-based methods in a unified experimental setting, we uncover limitations that are not apparent in within-paradigm evaluations, including the relative underperformance of LLM-based methods (RQ1) and their sensitivity to evaluation settings (RQ3). For researchers, this suggests that evaluations confined to a single paradigm may provide an incomplete or even misleading picture of method effectiveness. Future work should therefore incorporate cross-paradigm comparisons to better position new approaches against established alternatives. More broadly, this finding points to a general lesson for SE: meaningful evaluation requires not only strong baselines, but also comparisons across fundamentally different solution paradigms.

\section{Threats to Validity}
This section discusses potential threats to the validity of our empirical study.

\noindent\textbf{Selection of datasets.} The choice of datasets may threaten the validity of our results. To mitigate this threat, we use three widely studied real-world datasets spanning multiple high-stakes domains and a diverse range of sensitive attributes~\cite{hort2024bias, mehrabi2021survey, chen2022maat, chen2024fairness}. This setup provides broad coverage of application settings that are central to fairness research and enables a controlled comparison between the ML and LLM paradigms.

\noindent\textbf{Selection of mitigation methods.} Our results may also be affected by the mitigation methods included in the study. To reduce this threat, we include representative traditional ML methods spanning pre-processing, in-processing, post-processing, and ensemble learning, as well as representative LLM-based mitigation methods under in-context learning, which are widely adopted in prior studies~\cite{hort2024bias,ltdd, xiao2024mirrorfair, chen2025software, mehrabi2021survey, hu2024strategic, wang2023decodingtrust}. For RQ4, we additionally construct fine-tuning-based methods to examine whether exposing LLMs to the full training data changes the overall conclusion. These methods cover the main stages of bias mitigation in tabular classification.

\noindent\textbf{Selection of models.} Another threat lies in the choice of models. To mitigate this threat, we evaluate four widely used ML classifiers and use GPT-4o-mini as the primary under-test LLM. We also validate the main findings on four additional advanced LLMs in RQ2 and examine fine-tuning in RQ4 with two lightweight and economical LLMs. This design reduces the risk that our findings are tied to a single model family and allows us to assess whether the main conclusion remains stable across different model choices.

\noindent\textbf{Selection of evaluation settings and metrics.} The evaluation settings and metrics used in this study may also influence the results. To alleviate this threat, we align datasets, metrics, and evaluation settings across paradigms, and we explicitly compare balanced-test and random-test settings in RQ3. In addition, we adopt widely used~\cite{chen2025software,hort2024bias,diversityICSE} metrics for both predictive performance and fairness, and report both types of metrics instead of relying on a single criterion.

\noindent\textbf{Data leakage in LLMs.} The use of both closed-source and open-source LLMs introduces a threat related to potential data leakage from public data sources~\cite{li2024contamination, hidayat2025simulating, wu2024condefects}. To mitigate this threat, we evaluate multiple advanced LLMs in RQ2 and report repeated results rather than single runs. Although we cannot fully rule out data leakage, such leakage would be more likely to favor the LLM paradigm than disadvantage it. Despite this possibility, the zero-shot and few-shot LLM results in our study still remain clearly below traditional ML in many settings.

\section{Conclusion}
This paper presents a large-scale empirical comparison of state-of-the-art ML- and LLM-based bias mitigation methods for tabular classification. Our results show that traditional ML-based methods consistently outperform advanced LLM-based methods in both predictive performance and fairness, while the favorable fairness reported in prior LLM-based studies is highly sensitive to evaluation settings. Although supervised fine-tuning improves over in-context learning, it does not consistently outperform traditional ML-based methods. These findings suggest that LLM-based methods should not be treated as a default choice for fairness-critical tabular applications. They also indicate that the advantages of LLM-based methods over traditional approaches may not be consistent in broader SE tasks. More broadly, our study highlights that adopting emerging techniques in SE requires rigorous, cross-paradigm evaluation.

\section{Data Availability}
We have publicly released the scripts, data, and our analysis results as a replication package~\cite{githublink}.

\balance

\bibliographystyle{ACM-Reference-Format}
\bibliography{fairness}

@inproceedings{mahmoud2019performance,
  title={Performance predicting in hiring process and performance appraisals using machine learning},
  author={Mahmoud, Ali A and Shawabkeh, Tahani AL and Salameh, Walid A and Al Amro, Ibrahim},
  booktitle={2019 10th international conference on information and communication systems (ICICS)},
  pages={110--115},
  year={2019},
  organization={IEEE}
}

@article{abs240902977,
  author       = {Junwei Liu and
                  Kaixin Wang and
                  Yixuan Chen and
                  Xin Peng and
                  Zhenpeng Chen and
                  Lingming Zhang and
                  Yiling Lou},
  title        = {Large Language Model-Based Agents for Software Engineering: {A} Survey},
  journal      = {ACM Transactions on Software Engineering and Methodology},
  year         = {2026}
}

@article{donohue2018replacement,
  title={A replacement for Justitia's scales: Machine learning's role in sentencing},
  author={Donohue, Michael E},
  journal={Harv. JL \& Tech.},
  volume={32},
  pages={657},
  year={2018},
  publisher={HeinOnline}
}

@misc{credit_dataset,
    title = {The Credit Dataset},
    howpublished = {\url{https://archive.ics.uci.edu/dataset/350/default+of+credit+card+clients}},
    year = {1994}
}

@misc{adult_dataset,
    title = {The Adult Census Income dataset},
    howpublished = {\url{https://archive.ics.uci.edu/ml/datasets/adult}},
    year = {2017}
}

@misc{compas_dataset,
  title = {The Compas dataset},
  howpublished = {\url{https://github.com/propublica/compas-analysis}},
  year = {2016}
}

@misc{gpt4omini,
  title = {GPT-4o mini},
  howpublished = {\url{https://platform.openai.com/docs/models/gpt-4o-mini}},
  year = {2024}
}

@misc{qwen2.5,
  title = {Qwen2.5-7B-Instruct},
  howpublished = {\url{https://huggingface.co/Qwen/Qwen2.5-7B-Instruct}},
  year = {2024}
}

@misc{gpt5,
  title = {GPT-5},
  howpublished = {\url{https://developers.openai.com/api/docs/models/gpt-5-chat-latest}},
  year = {2025}
}

@misc{gemini2.5,
  title = {Gemini-2.5-Flash},
  howpublished = {\url{https://ai.google.dev/gemini-api/docs/models/gemini-2.5-flash}},
  year = {2025}
}

@misc{deepseek3.2,
  title = {DeepSeek-v3.2},
  howpublished = {\url{https://huggingface.co/deepseek-ai/DeepSeek-V3.2}},
  year = {2025}
}

@misc{openai,
  title = {openai},
  howpublished = {\url{https://developers.openai.com/api/reference/overview}},
  year = {2026}
}

@misc{openrouter,
  title = {openrouter},
  howpublished = {\url{https://openrouter.ai/}},
  year = {2026}
}

@misc{qwen3max,
  title = {Qwen3-Max-Thinking},
  howpublished = {\url{https://openrouter.ai/qwen/qwen3-max-thinking}},
  year = {2026}
}

@misc{aif360,
  title = {IBM AIF360},
  howpublished = {\url{https://ai-fairness-360.org/}},
  year = {2024}
}

@inproceedings{zhang2021ignorance,
  title={" Ignorance and Prejudice" in Software Fairness},
  author={Zhang, Jie M and Harman, Mark},
  booktitle={2021 IEEE/ACM 43rd International Conference on Software Engineering (ICSE)},
  pages={1436--1447},
  year={2021},
  organization={IEEE}
}

@article{chen2023comprehensive,
  title={A comprehensive empirical study of bias mitigation methods for machine learning classifiers},
  author={Chen, Zhenpeng and Zhang, Jie M and Sarro, Federica and Harman, Mark},
  journal={ACM transactions on software engineering and methodology},
  volume={32},
  number={4},
  pages={1--30},
  year={2023},
  publisher={ACM New York, NY, USA}
}

@article{hort2024bias,
  title={Bias mitigation for machine learning classifiers: A comprehensive survey},
  author={Hort, Max and Chen, Zhenpeng and Zhang, Jie M and Harman, Mark and Sarro, Federica},
  journal={ACM Journal on Responsible Computing},
  volume={1},
  number={2},
  pages={1--52},
  year={2024},
  publisher={ACM New York, NY}
}

@inproceedings{chen2022maat,
  title={MAAT: a novel ensemble approach to addressing fairness and performance bugs for machine learning software},
  author={Chen, Zhenpeng and Zhang, Jie M and Sarro, Federica and Harman, Mark},
  booktitle={Proceedings of the 30th ACM joint european software engineering conference and symposium on the foundations of software engineering},
  pages={1122--1134},
  year={2022}
}

@inproceedings{li2026fairness,
  author       = {Xinyue Li and
                  Zhenpeng Chen and
                  Jie M. Zhang and
                  Ying Xiao and
                  Tianlin Li and
                  Weisong Sun and
                  Yang Liu and
                  Yiling Lou and
                  Xuanzhe Liu},
  title        = {Fairness Testing of Large Language Models in Role-Playing},
  booktitle    = {Proceedings of the 34th {ACM} International Conference on the Foundations
                  of Software Engineering, {FSE}},
  year         = {2026}
}

@article{xiao2024mirrorfair,
  title={MirrorFair: Fixing fairness bugs in machine learning software via counterfactual predictions},
  author={Xiao, Ying and Zhang, Jie M and Liu, Yepang and Mousavi, Mohammad Reza and Liu, Sicen and Xue, Dingyuan},
  journal={Proceedings of the ACM on Software Engineering},
  volume={1},
  number={FSE},
  pages={2121--2143},
  year={2024},
  publisher={ACM New York, NY, USA}
}

@inproceedings{ltdd,
  title={Training data debugging for the fairness of machine learning software},
  author={Li, Yanhui and Meng, Linghan and Chen, Lin and Yu, Li and Wu, Di and Zhou, Yuming and Xu, Baowen},
  booktitle={Proceedings of the 44th International Conference on Software Engineering},
  pages={2215--2227},
  year={2022}
}

@inproceedings{zhang2018mitigating,
  title={Mitigating unwanted biases with adversarial learning},
  author={Zhang, Brian Hu and Lemoine, Blake and Mitchell, Margaret},
  booktitle={Proceedings of the 2018 AAAI/ACM Conference on AI, Ethics, and Society},
  pages={335--340},
  year={2018}
}

@inproceedings{chakraborty2020fairway,
  title={Fairway: a way to build fair ML software},
  author={Chakraborty, Joymallya and Majumder, Suvodeep and Yu, Zhe and Menzies, Tim},
  booktitle={Proceedings of the 28th ACM joint meeting on European software engineering conference and symposium on the foundations of software engineering},
  pages={654--665},
  year={2020}
}

@article{mehrabi2021survey,
  title={A survey on bias and fairness in machine learning},
  author={Mehrabi, Ninareh and Morstatter, Fred and Saxena, Nripsuta and Lerman, Kristina and Galstyan, Aram},
  journal={ACM computing surveys (CSUR)},
  volume={54},
  number={6},
  pages={1--35},
  year={2021},
  publisher={ACM New York, NY, USA}
}

@article{wies2023learnability,
  title={The learnability of in-context learning},
  author={Wies, Noam and Levine, Yoav and Shashua, Amnon},
  journal={Advances in Neural Information Processing Systems},
  volume={36},
  pages={36637--36651},
  year={2023}
}

@inproceedings{hu2024strategic,
  title={Strategic demonstration selection for improved fairness in llm in-context learning},
  author={Hu, Jingyu and Liu, Weiru and Du, Mengnan},
  booktitle={Proceedings of the 2024 Conference on Empirical Methods in Natural Language Processing},
  pages={7460--7475},
  year={2024}
}

@inproceedings{liu2024confronting,
  title={Confronting LLMs with traditional ML: Rethinking the fairness of large language models in tabular classifications},
  author={Liu, Yanchen and Gautam, Srishti and Ma, Jiaqi and Lakkaraju, Himabindu},
  booktitle={Proceedings of the 2024 Conference of the North American Chapter of the Association for Computational Linguistics: Human Language Technologies (Volume 1: Long Papers)},
  pages={3603--3620},
  year={2024}
}

@article{wang2023decodingtrust,
  title={DecodingTrust: A Comprehensive Assessment of Trustworthiness in $\{$GPT$\}$ Models},
  author={Wang, Boxin and Chen, Weixin and Pei, Hengzhi and Xie, Chulin and Kang, Mintong and Zhang, Chenhui and Xu, Chejian and Xiong, Zidi and Dutta, Ritik and Schaeffer, Rylan and others},
  year={2023},
  publisher={Neural Information Processing Systems Datasets; Benchmarks Track}
}

@inproceedings{li2024contamination,
  title={An open-source data contamination report for large language models},
  author={Li, Yucheng and Guo, Yunhao and Guerin, Frank and Lin, Chenghua},
  booktitle={Findings of the Association for Computational Linguistics: EMNLP 2024},
  pages={528--541},
  year={2024}
}

@article{chen2024fairness,
  title={Fairness testing: A comprehensive survey and analysis of trends},
  author={Chen, Zhenpeng and Zhang, Jie M and Hort, Max and Harman, Mark and Sarro, Federica},
  journal={ACM Transactions on Software Engineering and Methodology},
  volume={33},
  number={5},
  pages={1--59},
  year={2024},
  publisher={ACM New York, NY}
}

@article{soremekun2025software,
  title={Software fairness: An analysis and survey},
  author={Soremekun, Ezekiel and Papadakis, Mike and Cordy, Maxime and Le Traon, Yves},
  journal={ACM Computing Surveys},
  volume={58},
  number={3},
  pages={1--38},
  year={2025},
  publisher={ACM New York, NY}
}

@inproceedings{cherepanova2025improving,
  title={Improving llm group fairness on tabular data via in-context learning},
  author={Cherepanova, Valeriia and Lee, Chia-Jung and Akpinar, Nil-Jana and Fogliato, Riccardo and Lopez, Martin Bertran and Kearns, Michael and Zou, James},
  booktitle={Proceedings of the AAAI/ACM Conference on AI, Ethics, and Society},
  volume={8},
  number={1},
  pages={579--590},
  year={2025}
}

@inproceedings{biswas2020machine,
  title={Do the machine learning models on a crowd sourced platform exhibit bias? an empirical study on model fairness},
  author={Biswas, Sumon and Rajan, Hridesh},
  booktitle={Proceedings of the 28th ACM joint meeting on European software engineering conference and symposium on the foundations of software engineering},
  pages={642--653},
  year={2020}
}

@inproceedings{aydemir2018roadmap,
  title={A roadmap for ethics-aware software engineering},
  author={Aydemir, Fatma Ba{\c{s}}ak and Dalpiaz, Fabiano},
  booktitle={Proceedings of the international workshop on software fairness},
  pages={15--21},
  year={2018}
}

@inproceedings{brun2018software,
  title={Software fairness},
  author={Brun, Yuriy and Meliou, Alexandra},
  booktitle={Proceedings of the 2018 26th ACM joint meeting on european software engineering conference and symposium on the foundations of software engineering},
  pages={754--759},
  year={2018}
}

@inproceedings{diversityICSE,
title = {Diversity Drives Fairness: Ensemble of Higher Order Mutants for Intersectional Fairness of Machine Learning Software},
author = {Chen, Zhenpeng and Li, Xinyue and Zhang, Jie M. and Sarro, Federica and Liu, Yang},
booktitle = {Proceedings of the IEEE/ACM 47th International Conference on Software Engineering},
pages = {743–755},
year = {2025}
}

@inproceedings{fairsense,
title = {FairSense: Long-Term Fairness Analysis of ML-Enabled Systems},
author = {She, Yining and Biswas, Sumon and K\"{a}stner, Christian and Kang, Eunsuk},
booktitle = {Proceedings of the IEEE/ACM 47th International Conference on Software Engineering},
pages = {782–794},
year = {2025}
}

@article{peng2022fairmask,
  title={Fairmask: Better fairness via model-based rebalancing of protected attributes},
  author={Peng, Kewen and Chakraborty, Joymallya and Menzies, Tim},
  journal={IEEE Transactions on Software Engineering},
  volume={49},
  number={4},
  pages={2426--2439},
  year={2022},
  publisher={IEEE}
}

@article{cot,
  title={Fairness Is Not Just Ethical: Performance Trade-Off via Data Correlation Tuning to Mitigate Bias in ML Software},
  author={Xiao, Ying and Wang, Shangwen and Liu, Sicen and Xue, Dingyuan and Zhan, Xian and Liu, Yepang and Zhang, Jie M},
  journal={arXiv preprint arXiv:2512.21348},
  year={2025}
}

@article{kamiran2012data,
  title={Data preprocessing techniques for classification without discrimination},
  author={Kamiran, Faisal and Calders, Toon},
  journal={Knowledge and information systems},
  volume={33},
  number={1},
  pages={1--33},
  year={2012},
  publisher={Springer}
}

@article{eop,
  title={Equality of opportunity in supervised learning},
  author={Hardt, Moritz and Price, Eric and Srebro, Nati},
  journal={Advances in neural information processing systems},
  volume={29},
  year={2016}
}

@inproceedings{roc,
  title={Decision theory for discrimination-aware classification},
  author={Kamiran, Faisal and Karim, Asim and Zhang, Xiangliang},
  booktitle={2012 IEEE 12th international conference on data mining},
  pages={924--929},
  year={2012},
  organization={IEEE}
}

@inproceedings{shaikh2023second,
  title={On second thought, let’s not think step by step! bias and toxicity in zero-shot reasoning},
  author={Shaikh, Omar and Zhang, Hongxin and Held, William and Bernstein, Michael and Yang, Diyi},
  booktitle={Proceedings of the 61st Annual Meeting of the Association for Computational Linguistics (Volume 1: Long Papers)},
  pages={4454--4470},
  year={2023}
}

@article{DBLP:journals/corr/abs-2303-18223,
  author       = {Wayne Xin Zhao and
                  Kun Zhou and
                  Junyi Li and
                  Tianyi Tang and
                  Xiaolei Wang and
                  Yupeng Hou and
                  Yingqian Min and
                  Beichen Zhang and
                  Junjie Zhang and
                  Zican Dong and
                  Yifan Du and
                  Chen Yang and
                  Yushuo Chen and
                  Zhipeng Chen and
                  Jinhao Jiang and
                  Ruiyang Ren and
                  Yifan Li and
                  Xinyu Tang and
                  Zikang Liu and
                  Peiyu Liu and
                  Jian{-}Yun Nie and
                  Ji{-}Rong Wen},
  title        = {A survey of large language models},
  journal      = {CoRR},
  volume       = {abs/2303.18223},
  year         = {2023},
  eprinttype    = {arXiv}
}

@inproceedings{ChiangZ0ALLZ0JG24,
  author       = {Wei{-}Lin Chiang and
                  Lianmin Zheng and
                  Ying Sheng and
                  Anastasios Nikolas Angelopoulos and
                  Tianle Li and
                  Dacheng Li and
                  Banghua Zhu and
                  Hao Zhang and
                  Michael I. Jordan and
                  Joseph E. Gonzalez and
                  Ion Stoica},
  title        = {Chatbot arena: An open platform for evaluating LLMs by human preference},
  booktitle    = {Proceedings of the Forty-first International Conference on Machine Learning, {ICML} 2024},
  year         = {2024}
}

@inproceedings{biswas2021fair,
  title={Fair preprocessing: towards understanding compositional fairness of data transformers in machine learning pipeline},
  author={Biswas, Sumon and Rajan, Hridesh},
  booktitle={Proceedings of the 29th ACM joint meeting on European software engineering conference and symposium on the foundations of software engineering},
  pages={981--993},
  year={2021}
}

@inproceedings{chakraborty2021bias,
  title={Bias in machine learning software: Why? how? what to do?},
  author={Chakraborty, Joymallya and Majumder, Suvodeep and Menzies, Tim},
  booktitle={Proceedings of the 29th ACM joint meeting on european software engineering conference and symposium on the foundations of software engineering},
  pages={429--440},
  year={2021}
}

@article{mann1947test,
  title={On a test of whether one of two random variables is stochastically larger than the other},
  author={Mann, Henry B and Whitney, Donald R},
  journal={The annals of mathematical statistics},
  pages={50--60},
  year={1947},
  publisher={JSTOR}
}

@inproceedings{hidayat2025simulating,
  title={Simulating training data leakage in multiple-choice benchmarks for llm evaluation},
  author={Hidayat, Naila Shafirni and Al Kautsar, Muhammad Dehan and Wicaksono, Alfan Farizki and Koto, Fajri},
  booktitle={Proceedings of the 5th Workshop on Evaluation and Comparison of NLP Systems},
  pages={21--39},
  year={2025}
}

@inproceedings{wu2024condefects,
  title={Condefects: A complementary dataset to address the data leakage concern for llm-based fault localization and program repair},
  author={Wu, Yonghao and Li, Zheng and Zhang, Jie M and Liu, Yong},
  booktitle={Companion Proceedings of the 32nd ACM International Conference on the Foundations of Software Engineering},
  pages={642--646},
  year={2024}
}

@inproceedings{alidoosti2021ethics,
  title={Ethics-driven software architecture decision making},
  author={Alidoosti, Razieh},
  booktitle={2021 IEEE 18th International Conference on Software Architecture Companion (ICSA-C)},
  pages={90--91},
  year={2021},
  organization={IEEE}
}

@article{wick2019unlocking,
  title={Unlocking fairness: a trade-off revisited},
  author={Wick, Michael and Tristan, Jean-Baptiste and others},
  journal={Advances in neural information processing systems},
  volume={32},
  year={2019}
}

@article{protection2018general,
  title={General data protection regulation},
  author={Protection, Data},
  journal={Intersoft Consulting, Accessed in October},
  volume={24},
  number={1},
  year={2018}
}

@article{chen2025software,
  title={Software Fairness Dilemma: Is Bias Mitigation a Zero-Sum Game?},
  author={Chen, Zhenpeng and Li, Xinyue and Zhang, Jie M and Sun, Weisong and Xiao, Ying and Li, Tianlin and Lou, Yiling and Liu, Yang},
  journal={Proceedings of the ACM on Software Engineering},
  volume={2},
  number={FSE},
  pages={1780--1801},
  year={2025},
  publisher={ACM New York, NY, USA}
}

@misc{githublink,
  title = {Replication package},
  howpublished = {\url{https://doi.org/10.5281/zenodo.19244975}},
  year = {2026}
}

@inproceedings{chen2025more,
  title={More Women, Same Stereotypes: Unpacking the Gender Bias Paradox in Large Language Models},
  author={Chen, Evan and Zhan, Run-Jun and Lin, Yan-Bai and Chen, Hung-Hsuan},
  booktitle={Proceedings of the 34th ACM International Conference on Information and Knowledge Management},
  pages={4639--4643},
  year={2025}
}

@inproceedings{gorti2024unboxing,
  title={Unboxing occupational bias: Debiasing llms with us labor data},
  author={Gorti, Atmika and Chadha, Aman and Gaur, Manas},
  booktitle={Proceedings of the AAAI Symposium Series},
  volume={4},
  number={1},
  pages={48--55},
  year={2024}
}

@article{liu2025llms,
  title={Do LLMs Align Human Values Regarding Social Biases? Judging and Explaining Social Biases with LLMs},
  author={Liu, Yang and Chu, Chenhui},
  journal={arXiv preprint arXiv:2509.13869},
  year={2025}
}

@String{Computing = "Computing" }

@String{Springer = "Springer-Verlag" }

\end{document}